\documentclass[aps,groupedaddress]{revtex4-2}

\usepackage[a-3u]{pdfx}         
\usepackage{textcomp}           
\usepackage[T1]{fontenc}
\usepackage{emptypage}  
\usepackage{datetime}
\newdateformat{monthyeardate}{\monthname[\THEMONTH] \THEYEAR}

\makeatletter
\renewcommand{\p@subsection}{}
\renewcommand{\p@subsubsection}{}
\makeatother

\usepackage{url}                
\usepackage{hyperref}
\colorlet{benlinkcol}{black}
\colorlet{bencitecol}{black}
\colorlet{benurlcol}{black}
\hypersetup{
    pdftitle = {\Title},
    pdfauthor = {Benjamin Levi Cookman},
    pdflang = {en-GB},
    colorlinks = true,
    linkcolor = benlinkcol,
    citecolor = bencitecol,
    urlcolor = benurlcol
}

\usepackage{natbib}

\usepackage{physics}
\usepackage{amsmath}            
\allowdisplaybreaks[1]          
\usepackage{amssymb}            
\usepackage{amsbsy}             
\usepackage{siunitx}                    
\usepackage{mathtools}
\usepackage{bbm}
\usepackage{empheq}
\usepackage{gensymb}
\usepackage{esvect}                     
\usepackage{stmaryrd}
\usepackage{plimsoll}
\usepackage{MnSymbol}

\allowdisplaybreaks

\numberwithin{equation}{section}

\usepackage{rotating}           
\usepackage{graphicx,psfrag}    
\usepackage{multirow}           
\usepackage{tabularx}           
\usepackage{booktabs}                   
\usepackage{subcaption}                 
\usepackage{ragged2e}                   
\usepackage{array}              
\usepackage[font=small]{caption}            
\usepackage{enumitem}
\usepackage{scalerel}           
\usepackage{accents}
\usepackage{footnote}           
\usepackage[bottom]{footmisc}   
\usepackage{fancyhdr}           
\usepackage{verbatim}                   
\usepackage{lipsum}                     
\usepackage[base]{babel}                
\usepackage{stackengine}
\usepackage[super]{nth}

\newcommand*\widefbox[1]{\fbox{\hspace{1.5mm}#1\hspace{1.5mm}}}
\stackMath
\newcommand\widecheck[1]{%
\savestack{\tmpbox}{\stretchto{%
  \scaleto{%
    \scalerel*[\widthof{\ensuremath{#1}}]{\kern-.6pt\bigwedge\kern-.6pt}%
    {\rule[-\textheight/2]{1ex}{\textheight}}   
  }{\textheight}%
}{0.5ex}}%
\stackon[1pt]{\displaystyle #1}{\scalebox{-1}{\tmpbox}}%
}
\renewcommand\widehat[1]{%
\savestack{\tmpbox}{\stretchto{%
  \scaleto{%
    \scalerel*[\widthof{\ensuremath{#1}}]{\kern-.6pt\bigvee\kern-.6pt}%
    {\rule[-\textheight/2]{1ex}{\textheight}}   
  }{\textheight}%
}{0.5ex}}%
\stackon[1pt]{\displaystyle #1}{\scalebox{-1}{\tmpbox}}%
}

\newcommand{\sus}[1]{$^{\mbox{\scriptsize #1}}$}

\newcommand{\sect}[1]{Section~\ref{#1}}
\newcommand{\fig}[1]{Fig.~\ref{#1}}
\newcommand{\equ}[1]{(\ref{#1})}
\newcommand{\appx}[1]{Appendix~\ref{#1}}

\newcommand{\g}{\gamma}

\renewcommand{\d}{\delta}
\newcommand{\D}{\Delta}

\renewcommand{\k}{\kappa}
\renewcommand{\l}{\lambda}
\renewcommand{\L}{\Lambda}
\newcommand{\h}{\eta}
\newcommand{\m}{\mu}
\renewcommand{\r}{\rho}
\newcommand{\p}{\pi}

\renewcommand{\t}{\tau}
\newcommand{\s}{\sigma}

\renewcommand{\r}{\rho}
\renewcommand{\o}{\omega}

\newcommand{\f}{\varphi}

\renewcommand{\rm}{\mathrm}
\newcommand{\bb}[1]{\mathbb{#1}}
\newcommand{\cl}[1]{\mathcal{#1}}

\newcommand{\Le}{\rm{Le}}
\renewcommand{\Pr}{\rm{Pr}}
\newcommand{\Ma}{\rm{Ma}}
\newcommand{\Ze}{\rm{Ze}}
\newcommand{\Mk}{\rm{M}_{\rm{k}}}

\renewcommand{\vb}[1]{\boldsymbol{#1}}

\newcommand{\ftvar}[1]{\tilde{#1}}

\newcommand{\und}[1]{\vb{#1}}
\newcommand{\dimvar}[1]{#1^*}

\newcommand{\vnab}{\vb{\nabla}}

\newcommand{\Mod}[1]{\ (\mathrm{mod}\ #1)}

\newcommand*{\tran}{^{\mkern-1.5mu\mathsf{T}}}

\newcommand{\Title}{Direct Numerical Simulation of Thermoacoustically Unstable Flames Via Non-Drifting Acoustic Delay Characteristic Boundary Conditions}

\begin{document}

\title{\Title}
\author{Benjamin L. Cookman}
\email{benjamin.cookman@manchester.ac.uk}
\affiliation{School of Engineering, University of Manchester}
\author{Jack R. C. King}
\affiliation{School of Engineering, University of Manchester}
\author{Raphaël C. Assier}
\affiliation{Department of Mathematics, University of Manchester}
\author{Steven J. Lind}
\affiliation{School of Engineering, Cardiff University}
\date{November 2026}

\begin{abstract}

In this work, Direct Numerical Simulations (DNS) of thermoacoustically unstable free flames in closed-open tubes are presented. Simulations demonstrate nonlinear saturation of the primary instability as the flame flattens, resulting in a quasiperiodic acoustic limit cycle. Acoustic forcing is used to trigger secondary parametric instability, causing an oscillatory flame fingering characteristic of the unsteady Rayleigh-Taylor effect. To reduce computational cost, we perform DNS only on the region surrounding the flame. Acoustics at in- and outflows are described using the Navier-Stokes Characteristic Boundary Condition (NSCBC) method to model their delayed reentry into the domain in a formulation referred to as the Acoustic Delay Characteristic Boundary Condition (ADCBC) method. A new Averaged Proportional and Integral Linear Relaxation (APILR) method is also introduced, which modifies the Classic Linear Relaxation (CLR) method to maintain time-averaged values of inflow velocity and outflow pressure. Many moving averages may be used consecutively to increase the damping rate of boundary reflection coefficients. An integral control term is used to remove non-zero equilibrium time-averaged inflow velocities which adversely affect control over flame position in the DNS region. Both new boundary methods demonstrate their capability in inert and counterflow flames test cases. In the case of a 100~cm long tube, these methods enable the numerical simulation of combustion instabilities at 98\% reduced computational expense.

\end{abstract}

\maketitle

\section{Introduction}
Fuel-deficient, or lean, combustion presents an attractive way of avoiding the harmful NO$_x$ emissions produced via high-temperature combustion with nitrogen in air. But, this renders the flow more susceptible to thermoacoustic instabilities~\cite{mongia2003ChallengesProgressControlling}. Thermoacoustic instabilities are caused by the coupling of a flame's unsteady heat release with the surrounding acoustic field. Provided this coupling is in phase, it was predicted by Rayleigh~\cite{rayleigh1878TheorySound} that this may result in positive feedback. This was later verified as Mallard and Le Chatellier observed acoustic waves emanating from confined flames~\cite{mallard1883RecherchesExperimentalesTheoriques}. Both intrinsic (geometry independent) and extrinsic (geometry dependent) modes of thermoacoustic instability exist; we aim the focus of this paper at extrinsic modes, as these remain challenging to model and predict despite sustained research in the area~\cite{juniper2018SensitivityNonlinearityThermoacoustic,morgans2024ThermoacousticInstabilityCombustors}. In this case, thermal-acoustic coupling occurs due to the reflection of acoustic waves away from the flame at the combustion chamber's boundaries, which correspond to perturbations of the whole system's acoustic modes~\cite{morgans2024ThermoacousticInstabilityCombustors}.

Two types of acoustic combustion instability exist. Under primary instability, thermal-acoustic coupling results in a linear instability known to stabilise the intrinsically unstable flame front, causing a reduction in the growth rate of the acoustic modes~\cite{searby1992AcousticInstabilityPremixed}. Secondary instability is only observed once an acoustic amplitude threshold is exceeded and results in rapid growth of flame curvature and acoustic amplitudes. When a threshold acoustic amplitude is met, parametric instability occurs where higher wavenumber perturbations to the flame front grow proportional to the forcing frequency. A cellular flame front is formed which oscillates with double the period of the dominant acoustic mode (subharmonic oscillation). This results in a rapid increase in flame surface area, increasing the speed of the flame and Reynolds number of the flow. This typically renders the flame susceptible to hydrodynamic instabilities which lead to a turbulent flame. Further high-quality imagery of each stage of instability is presented in~\cite{dubey2019EffectGeometricalParameters,delfin2024VideoTransientParametric}. The presence of either mode can result in structural wear-and-tear to the combustion chamber and the drastic change to flame structure posed by parametric instability can rapidly cause damage. Another obvious downside is that of noise pollution: in jet engines, for example, as non-combustion noise sources have been reduced, combustion noise has become a proportionally higher contributor to overall noise~\cite{dowling2015CombustionNoise}.

Experimental studies form the bulk of research performed into combustion instability, as physical mechanisms are directly enquired. A major complexity when modelling these instabilities, is the three-way multiscale coupling between the flame, surrounding hydrodynamics and acoustics of the combustion chamber. The coupled flame-flow system is subject to intrinsic flame instabilities which affect thermoacoustic modes, such as Darrieus-Landau (DL)~\cite{darrieus1945PropagationDunFront,landau1944TheorySlowCombustion,matalon2018DarrieusLandauInstability} and thermodiffusive instability~\cite{zeldovich1944TheoryCombustionDetonation,barenblatt1962DiffusionalThermalStabilityLaminar,sivashinsky1977DiffusionalThermalTheoryCellular}. Various theoretical models exist for these two thermoacoustic modes, but they remain idealised and limited in scope. Using matched asymptotic expansions, it is possible to rigorously involve all first order interactions in each of the three physical scales~\cite{assier2014LinearWeaklyNonlinear}. By describing the acoustics as an oscillating body force, an unsteady Rayleigh-Taylor (RT) effect on the flame surface is revealed. However, this relies on the flame being described by a function in the transverse duct dimension, which disallows the more complex nonlinear parametric flame structures observed in experimentation (e.g.~in~\cite{searby1992AcousticInstabilityPremixed,clavin2016CombustionWavesFronts,delfin2024ThermoacousticParametricInstability}) and which is expected under the nonlinear RT instability. The Mathieu equation model, first introduced phenomenalogically in~\cite{markstein1951ExperimentalTheoreticalStudies} and extended by~\cite{searby1986WeaklyTurbulentWrinkled,searby1991ParametricAcousticInstability} describes the flame subjected to acoustic forcing as decoupled damped harmonic oscillators for each wavenumber flame perturbation. This incorporates acoustic background as well as intrinsic instabilities. An alternative formulation which allows for more complex flame structures would be one which describes the flame as a level set. The so-called G-equation was introduced first by~\cite{markstein1964NonsteadyFlamePropagation} and describes flame motion kinematically according to its speed in the normal direction. In~\cite{dowling1999KinematicModelDucted}, for example, heat release fluctuations of a ducted flame are coupled to upstream flow fluctuations via a flame transfer function depending on the frequency of incident perturbations. Such a transfer function is also commonly used in network models, such as those used in~\cite{noiray2008UnifiedFrameworkNonlinear}. Therein, an additional dependence on upstream perturbation amplitude is allowed in their flame describing function to fully characterise the nonlinear flame response to acoustics of varying frequencies and amplitudes.

As a complement to expertimentation and theory, numerical simulations of these flows can be performed. Direct Numerical Simulations (DNS) solve the governing equations without extra modelling via a resolved discretisation of the problem in space~\cite{orszag1970AnalyticalTheoriesTurbulence} and time, providing all the complex nonlinear interaction described by the equations as allowed by the given discretisation. By comparison, Large Eddy Simulations (LES) use coarser meshes and approximate the sub-grid scale viscosity and flame interactions by means of closure models~\cite{yang2015LargeEddySimulationPresent, veynante2002TurbulentCombustionModeling}. Despite the relative accessibility of CFD codes, simulation data of acoustic combustion instabilities remain scarce. This is due to the computational challenge associated with resolving thin reactive and diffusive regions in space and simulating acoustics in the full domain, all whilst obeying hyperbolic and parabolic stability constraints on time step. On top of this, many popular combustion codes use the low-Mach assumption (such as PeleLM~\cite{amrex-combustionPeleLM}, HOLOMAC~\cite{motheau2016HighorderNumericalAlgorithm} Nek5000~\cite{nek1996NEK5000}), which precludes the relevant acoustic waves. In~\cite{gonzalez1996AcousticInstabilityPremixed}, a small rectangular domain is discretised for flame modelled by an idealised one-step reaction. The archetypal cellular parametric structure is observed, but the domain geometry is not representative of typical combustor scales. Since then, computational power has increased and CFD techniques have improved greatly. In~\cite{jun2023ParametricInstabilityPropagating} a 70 cm long, 20 mm diameter axisymmetric cylindrical tube is modelled for reacting methane-hydrogen-air mixtures. Subharmonic oscillation is observed without a cellular structure, which oscillates between a characteristic RT finger and tulip flame due to isothermal, no-slip wall boundaries (tulip flames are explored further in e.g.~\cite{ponizy2014TulipFlameMechanism}). However, due to the size of geometry which is discretised, approximately 200,000 core hours are required in each case simulated despite only second-order discretisations being made.

Being that combustor geometries are of utmost importance to acoustic behaviour, the effect of flame confinement has also been studied. The impact of wall boundaries on flame structures subject to thermoacoustic instability is studied numerically in~\cite{chen2026AcousticResponseAsymmetric}. These non-adiabatic, non-isothermal narrow channel promote a more complex flame structure with higher amplitude pressure oscillations despite thermal losses. In these numerical studies, downstream acoustic boundaries are almost unanimously considered as either perfectly or nearly perfectly reflecting (i.e. having constant downstream pressure) for modelling convenience. The impact of downstream exhaust dynamics on confined thermoacoustic instability was investigated in~\cite{rodriguez-gutierrez2026EffectInducedOuter}. Inclusion of this exhaust flow in an extended simulation domain results in large changes to flame structure as well as thermoacoustic and aeroacoustic dynamics.

In DNS of non-thermoacoustic combustion, non-reflecting boundary conditions are widely used to cut off the acoustic feedback at the computational boundary to inhibit acoustic combustion instabilities. These are most commonly implemented using the Navier-Stokes Characteristic Boundary Conditions (NSCBC) formulation, where at each boundary node the Locally One-Dimensional Inviscid (LODI) approximation is employed to separate incoming from outgoing characteristic waves of the reacting Navier-Stokes equations~\cite{thompson1990TimeDependentBoundaryConditions,poinsot1992BoundaryConditionsDirect}. To enforce non-reflecting conditions, acoustic waves entering the computational domain must vanish. These are combined with diffusive conditions at these boundaries to maintain well-posedness when diffusive terms are included in the governing equations~\cite{strikwerda1976InitialBoundaryValue,sutherland2003ImprovedBoundaryConditions}. Typically, Classic Linear Relaxation (CLR) is used in the NSCBC formulation to maintain either a target inflow velocity~\cite{poinsot2001TheoreticalNumericalCombustion} or outflow pressure~\cite{rudy1980NonreflectingOutflowBoundary,poinsot1992BoundaryConditionsDirect}. The CLR method takes the form of a Proportional (P) controller, but sacrifices the acoustic non-reflection condition for outgoing low frequency acoustics~\cite{selle2004ActualImpedanceNonreflecting}. To improve the response to low frequencies, it was suggested in~\cite{polifke2006PartiallyReflectingNonreflecting} that these waves be directly identified in the domain by a series of sample planes parallel to the boundary within the computational domain. This method is referred to as wave masking. Alternatively,~\cite{daviller2019GeneralizedNonreflectingInlet} identifies waves by integrating outgoing acoustic wave fluxes at the boundary. By including the sampled waves in the P controller, reflections vanish by construction. This formulation allows the authors to isolate the impact of acoustic forcing in a turbulent, slot-burner methane-air flame. It is noted in~\cite{dupuy2026LowreflectionFastconvergenceBoundary} that non-reflecting and non-drifting boundary conditions are caught in a compromise between two regimes, in that they must simultaneously: not reflect the physical acoustic waves expected to leave the domain and correct the relevant boundary variables from drifting over longer time scales. This correction, however, corresponds to applying a superposition of low frequency reflections where needed. Any resulting formulation is likely, then, to have close to full reflections for frequencies associated with relaxation time scales and diminishing reflections for frequencies corresponding to the physical system's acoustics. The work of~\cite{dupuy2026LowreflectionFastconvergenceBoundary} uses the same wave sampling formulation as~\cite{daviller2019GeneralizedNonreflectingInlet}, but improves upon convergence times by including a low-pass filter of the sampled outgoing acoustics in the P controller. The inclusion of this term spoils the lack of reflection at low frequencies, but improves on the rate of damping as frequency increases compared to CLR~\cite{selle2004ActualImpedanceNonreflecting}.

In this article we present a new boundary condition framework which significantly reduces the computational cost associated with the simulation of thermoacoustic instabilities by truncating the DNS domain. Currently, the limited understanding of thermoacoustic flame and flow behaviour \emph{a priori} make reducing simulation cost a challenge beyond use of adaptive CFD methods. In~\cite{douasbin2018AcousticWavesCombustion} the acoustic domain is truncated downstream of the domain by modelling the frequency dependence at the outflow boundary to match the response of a standing wave reflected downstream. This is implemented as an impedance boundary condition referred to as delayed Time-Domain Impedance Boundary Conditions (D-TDIBC). The truncated domain reproduces the correct acoustic modes surrounding a flame anchored to a cylinder in a tube at reduced computational expense. In this article, we truncate the acoustic domain up- and downstream of the flame in a similar manner. Acoustic waves in these truncated regions are modelled using the NSCBC formulation via their delayed reentry into the computational domain. This is referred to as the Acoustic Delay Characteristic Boundary Condition (ADCBC) method.

We also introduce a new drift control scheme compatible with these thermoacoustic systems which converges mean inflow velocities and outflow pressures to their target values whilst ignoring the acoustic modes. This scheme possesses reflection coefficients which vanish for these acoustic modes and which decreases at an arbitrary rate in decades per decade above an averaging frequency. The moving averages are reminiscient of~\cite{dupuy2026LowreflectionFastconvergenceBoundary}, except that we average the whole drift term repeatedly to occlude predetermined acoustic waves. Purely proportional control results in non-zero equilibrium drift values in the presence of a constant drift rate in the ADCBC method, so we also introduce an Integral term to converge exactly on target inflow velocity values. Hence, the premixed flame can be contained in the DNS region over longer simulation times. The resulting relaxation terms are referred to as the Averaged Proportional and Integral Linear Relaxation (APILR) method.

The ADCBC method enables DNS of the primary instability growth and flame flattening as well as the scarcely simulated cellular parametric flame instabilities. The focus of this work are developments in boundary conditions, so the results can be applied to any existing e.g.~finite difference or finite volumes based code making use of the NSCBC method. Herein, the high-order, meshless SUNSET combustion DNS code is used~\cite{king2024MeshFreeFrameworkHighOrdera}.

The paper is structured as follows. In \sect{sec:2} we introduce the NSCBC formulation, governing equations and CLR. In \sect{sec:3} we describe the delayed acoustics model and explain its application to the NSCBC formulation. We then apply the method to inert flows of travelling and standing acoustic waves to validate the ADCBC method. In \sect{sec:4} the new APILR relaxation terms are introduced to account for CLR being incompatible with these thermoacoustic problems when using the ADCBC method. In \sect{sec:5} we apply the ADCBC and APILR methods to premixed flame simulations of thermoacoustically unstable flames. In \sect{sec:6} we draw conclusions from the preceding sections.

\section{Navier-Stokes Characteristic Boundary Conditions} \label{sec:2}
\subsection{Governing Equations}

We solve the Navier-Stokes equations for a mixture of two species, a reactant R and product P, reacting irreversibly in a reaction R $\to$ P. Having said this, the results contained may also be extended trivially to mixtures of multiple species modelled under mixture averaged transport reacting in many steps. The governing equations for a three-dimensional flow $\vb{x} = (x, y, z)\tran$, $\vb{u} = (u, v, w)\tran$ are:
\begin{subequations} \label{eqn:EQUATIONS-DIFF}
\begin{alignat}{1}
\pdv{\r}{t} + \vnab \cdot(\r\vb{u}) &= 0, \\
\pdv{\r \vb{u}}{t} + \vnab  \cdot (\r \vb{u} \vb{u}\tran) &= -\vnab p + \vnab \cdot\mathbf{T} + \r \vb{g}, \\
\pdv{\r e}{t} + \vnab  \cdot (\r e \vb{u}) &= \l \D T - \D h_r^\plimsoll \vnab \cdot (\r D \vnab Y) + \vnab \cdot(\mathbf{S} \vb{u})+ \r \vb{g} \cdot \vb{u} + \r\dot{\cl{E}}, \\
\pdv{\r Y}{t} + \vnab \cdot (\r Y \! \vb{u}) &= \dot{\o} + \vnab \cdot(\r D \vnab Y),
\end{alignat}
\end{subequations}
alongside the equations for closure:
\begin{subequations} \label{eqn:EQUATIONS-ALGE}
\begin{align}
p &= \r \frac{R_0}{M} T, \\
\r e &= \r \D h_r^\plimsoll + \r \left( c_p - \frac{R_0}{M} \right) T + \frac{1}{2} \r \vb{u}\cdot\vb{u}.
\end{align}
\end{subequations}
Dependent variables are mass density $\r$, velocity $\vb{u}$, pressure $p$, specific internal and chemical energy $e$, temperature $T$ and product mass fraction $Y$. Reactant mass fraction is defined by mass conservation as $1 - Y$. The universal gas constant is denoted $R_0$. The terms $\r \vb{g}$ and $\r\dot{\cl{E}}$ are the body force and energy source terms. The tensors $\mathbf{S}$ and $\mathbf{T}$ are the usual stress and viscous stress tensors, respectively:
\begin{subequations}
\begin{align}
\mathbf{S} &:= -p \mathbf{I} + \mathbf{T}, \\
\mathbf{T} &:= \m  \left( - \frac{2}{3}(\vnab \cdot\vb{u}) \mathbf{I} + \vnab (\vb{u}\tran) + (\vnab (\vb{u}\tran))\tran \right)
\end{align}
\end{subequations}
where $\mathbf{I}$ is the $3 \times 3$ indentity matrix. Reaction rate is defined by the Arrhenius term:
\begin{equation}
\dot{\o} := A(1 - Y)\exp\left( - \frac{T}{R_0 E_a} \right).
\end{equation}

Besides initial and boundary conditions, the physical system is determined by the choice of constant fluid and reaction properties. The fluid properties are: molecular mass $M$, specific heat capacity $c_p$, viscosity $\m$, heat conducitivity $\l$ and molecular diffusion $D$. The reaction properties are: reaction enthalpy $\D h_r^\plimsoll$, pre-exponential factor $A$ and activation energy $E_a$. Soret and Dufour effects, pressure-gradient diffusion, and radiant heat flux effects are neglected.

\subsection{Characteristic Waves}

The Navier-Stokes Characteristic Boundary Condition (NSCBC) formulation approximates the fluid as Locally One-Dimensional and Inviscid (LODI), such that all diffusive and transverse terms are omitted. Excluding these terms in \equ{eqn:EQUATIONS-DIFF} gives us the Euler equations for a perfect reacting gas~\cite{poinsot2001TheoreticalNumericalCombustion} in terms of its six conserved variables ${\und{U} = (\r, \r \vb{u}, \r e, \r Y)\tran}$. The resulting hyperbolic system can then be decomposed into its constituent waves along a given dimension. Presuming the $x$-axis is normal to the boundary, we can make a choice of six convenient primitive variables ${\und{V} = (\r, \vb{u}, p, Y)\tran}$ and transform this into a system for $\und{V}$:
\begin{equation} \label{eqn:with_A}
\pdv{\und{V}}{t} + \mathbf{A}_x \pdv{\und{V}}{x} + \und{b} = \und{0}.
\end{equation}
The LODI assumption has been used to remove the transverse flux terms involving $\partial \und{V} / \partial y$ and $\partial \und{V} / \partial z$ and the vector $\und{b}$ only contains source terms acting on the primitive variables. The $6\times 6$ matrix $\mathbf{A}_x = \mathbf{A}_x(\und{V})$ represents nonlinear terms like the advection terms in the $x$-direction and has real eigenvalues $\l_m = \l_m(\und{V})$ and left-eigenvectors $\und{\ell}_m = \und{\ell}_m(\und{V})$ for $m = 1,\dots,6$. Multiplying \equ{eqn:with_A} by $\und{\ell}_m\tran$, we project onto a solution space containing only the $m$\sus{th} characteristic invariant~\cite{thompson1987LecturesSeriesComputational,thompson1987TimeDependentBoundary}, $J_m$, satisfying $\dd{J_m} = \und{\ell}_m\tran \dd{\und{V}} + \und{\ell}_m\tran \und{b} \dd{t}$:
\begin{equation} \label{eqn:single_char_prob}
\und{\ell}_m\tran \pdv{\und{V}}{t} + \cl{L}_m + \und{\ell}_m\tran \und{b}
= \pdv{J_m}{t} + \l_m \pdv{J_m}{x} = 0.
\end{equation}
where:
\begin{equation}
\cl{L}_m \equiv \l_m \und{\ell}_m\tran \pdv{\und{V}}{x}.
\end{equation}
Therefore, for each value of $m$, $J_m$ determines the value of the characteristic along a path
\begin{equation}
\vb{x}_m(t)=(x_m(t), y_m(t), z_m(t))\tran
\quad\text{with velocity}\quad
\l_m = \dv{x_m}{t}
\end{equation}
in the $x$-direction. Under the LODI formulation, waves $J_m$ which leave the computational domain through the boundary are determined by evaluating $\cl{L}_m$ via upwinded one-sided derivatives from within the computational domain. Characteristics entering the domain are determined by assigning $\cl{L}_m$ according to a physical model.

Solving the eigenvalue problem of $\mathbf{A}_x$ using the chosen primitive variables, we get $\l_1 = u-c$, $\l_5 = u+c$ and all other $\l_m = u$ with~\cite{poinsot2001TheoreticalNumericalCombustion}:
\begin{subequations} \label{eqn:λ_l_L}
\begin{alignat}{3}
\und{\ell}\tran_1 &= (0, -\r c, 0, 0, 1, 0), ~ &&  \cl{L}_1 &&= \l_1 \left(\pdv{p}{x} - \r c \pdv{u}{x}\right), \\
\und{\ell}\tran_2 &= (c^2, 0, 0, 0, -1, 0),  ~ &&  \cl{L}_2 &&= \l_2 \left(c^2 \pdv{\r}{x} - \pdv{p}{x}\right), \\
\und{\ell}\tran_3 &= (0, 0, 1, 0, 0, 0),     ~ &&  \cl{L}_3 &&= \l_3 \pdv{v}{x}, \\
\und{\ell}\tran_4 &= (0, 0, 0, 1, 0, 0),     ~ &&  \cl{L}_4 &&= \l_4 \pdv{w}{x}, \\
\und{\ell}\tran_5 &= (0, \r c, 0, 0, 1, 0),  ~ &&  \cl{L}_5 &&= \l_5 \left(\pdv{p}{x} + \r c \pdv{u}{x}\right), \\
\und{\ell}\tran_6 &= (0, 0, 0, 0, 0, 1),     ~ &&  \cl{L}_6 &&= \l_6 \pdv{Y}{x}.
\end{alignat}
\end{subequations}
where $c = \sqrt{\partial p / \partial \r}$ is the speed of sound. Hence, the characteristics $m = 1, 5$ correspond to acoustic waves travelling to the left and right along the $x$-axis, respectively. All other waves travel only with the fluid and correspond to advected waves. These are entropy waves ($m = 2$), vorticity waves ($m = 3, 4$) and species waves ($m = 6$). The equations of motion under these LODI assumptions become:
\begin{subequations}
\begin{alignat}{3}
& \pdv{\r}{t} &&+ \frac{1}{c^2} \left[ \cl{L}_2 + \frac{1}{2} \left( \cl{L}_5 + \cl{L}_1 \right) \right] &&= 0\\
& \pdv{u}{t}  &&+ \frac{1}{2 \r c}\left(\cl{L}_5 - \cl{L}_1\right) &&= g_x\\
& \pdv{v}{t}  &&+ \cl{L}_3 &&= g_y\\
& \pdv{w}{t}  &&+ \cl{L}_4 &&= g_z\\
& \pdv{p}{t}  &&+ \frac{1}{2}\left(\cl{L}_5 + \cl{L}_1 \right) &&= 0\\
& \pdv{Y}{t}  &&+ \cl{L}_6  &&= \dot{\o} / \r.
\end{alignat}
\end{subequations}
In this work, inflows are left-side boundaries and outflows are right-side boundaries (assuming acoustic waves of large enough amplitudes do not overcome the inflow or outflow velocity). We first identify the acoustic waves as either entering, $J_+$ or exiting, $J_-$ the domain. We define similarly $\cl{L}_+$ and $\cl{L}_-$. For inflows, the acoustic wave entering the domain is $J_+ = J_5$ and $\cl{L}_+ = \cl{L}_5$ and likewise for outflows $J_+ = J_1$ and $\cl{L}_+ = \cl{L}_1$ (and vice versa for waves exiting the domain). At inflows, the LODI strategy is to evaluate the outgoing $\cl{L}_-$ values via upwinded difference operators and model the incoming $\cl{L}_{+, 2, 3, 4, 6}$ values. At outflows, instead evaluate the outgoing $\cl{L}_{-, 2, 3, 4, 6}$ values via upwinded difference operators and model the incoming $\cl{L}_+$ values.

\subsection{Non-Reflecting Boundaries} \label{sec:nonreflect}

The NSCBC method is used frequently in the literature to impose non-reflection of acoustic waves at in- and outflow boundaries. We can define the reflection coefficient, $\hat{R}$ for a harmonic incident wave with frequency $f$ at this boundary as the ratio:
\begin{equation}
\hat{R}(f) := \frac{\hat{J}_+(f)}{\hat{J}_-(f)}.
\end{equation}
Variables with a hat $\hat{\cdot}$ are the Fourier transform of that variable in time. The resulting gain and phase of the boundary are $\abs{\hat{R}\,}(f)$ and $\arg(\hat{R})(f)$, respectively. Then, choosing $\cl{L}_+ = \r c g_x =: \cl{L}_+^{\rm{NR}}$ above implies $\hat{R} \equiv 0$. The remaining shear and species waves are determined by $\cl{L}_3$, $\cl{L}_4$ and $\cl{L}_6$. At inflows, the entrance of these waves is removed by assigning $\cl{L}_3^{\rm{NR}} = g_y$, $\cl{L}_4^{\rm{NR}} = g_z$ and $\cl{L}_4^{\rm{NR}} = \dot{\o} / \r$. Similarly, the remaining entropy waves $\cl{L}_2^{\rm{NR}}$ are assigned zero. In the remainder of the manuscript we assume no body force or reactions at the boundary, which requires vanishing non-reflection terms $\cl{L}_{2, 3, 4, 5, 6}^{\rm{NR}} \equiv 0$, e.g.~as used in~\cite{hedstrom1979NonreflectingBoundaryConditions,thompson1990TimeDependentBoundaryConditions,poinsot1992BoundaryConditionsDirect,poinsot2001TheoreticalNumericalCombustion}.

\subsection{Relaxation Terms}

The non-reflecting boundary conditions pose no condition on the desired in- and outflow pressure and velocity fields outside of initial conditions, allowing these fields to drift in the domain as the system remains open in an unphysical (and ill-posed) way. To complete the NSCBC formulation, drift is taken care of via CLR. These come in the form of extra terms on top of the existing non-reflecting ones:
\begin{equation} \label{eqn:clr}
\cl{L}_+ = \cl{L}_+^{\rm{NR}} + \cl{L}_+^{\rm{Relax}}
\end{equation}
where
\begin{equation}
\cl{L}_{+, \rm{IN}}^{\rm{Relax}}  := K s_{\rm{IN}}
\quad \text{and} \quad
\cl{L}_{+, \rm{OUT}}^{\rm{Relax}} := K s_{\rm{OUT}}.
\end{equation}
The coefficient $K$ has units of inverse time and determines the timescale of response for a given drift signal
\begin{equation}
s_{\rm{IN}} := \r c(u - u^\rm{t})
\quad \text{and} \quad
s_{\rm{OUT}} := (p - p^\rm{t})
\end{equation}
for target inflow velocity and outflow pressure values $u^\rm{t}$ and $p^\rm{t}$, respectively. Optimal values of $K$ for non-reflecting DNS are given by $K = 2\s / t_x$ where $t_x = 2L_x / c$ is the acoustic period of the computational domain and $\s \simeq 0.287$~\cite{rudy1981BoundaryConditionsSubsonic}. The resulting reflection coefficient becomes~\cite{selle2004ActualImpedanceNonreflecting}:
\begin{equation}
\hat{R}_{\rm{CLR}}(f) = \frac{1}{1 + i 2\p\!f t_x / \s}.
\end{equation}
As such, the boundaries behave as a fully reflecting boundary with $|\hat{R}_{\rm{CLR}}(f)| \approx 1$ for values of $f < f^c$ below a cutoff frequency $f^c := \s / t_x$ for a DNS domain with characteristic acoustic time scale $t_x$. The resulting reflection coefficient decreases as the incident frequency increases relative to the domains characteristic acoustic frequency.


\section{Acoustic Delay Model} \label{sec:3}
Investigating the physics of a thermoacoustic interaction means resolving the acoustics of the combustion system (this may involve a plenum, inlet, combustor and exhaust) the highly localised reaction and diffusion layers located at the flame and the surrounding gasdynamics. In the case of a deflagration in a tube of length $L_{\rm{tube}}$ and width $W$, the acoustics span the tube, but the premixed flame and flow are contained to an $\cl{O}(W)$ region surrounding the flame. For tubes where $W \ll L_{\rm{tube}}$, it would be natural to fully resolve the flame and flow in a simulated region and leave the acoustics up- and downstream of the flame to be modelled separately. We refer to these non-DNS domains as the acoustic domains. Consider a flame region of length $L_x$ and truncated acoustic domains up- and downstream. Each of these acoustic domains is characterised by its: length between the simulation boundary and the far acoustic boundary ($L_{\rm{U}}$ and $L_{\rm{D}}$, respectively), reflection coefficient at the far boundary ($R_{\rm{U}}$ and $R_{\rm{D}}$, respectively), flow speed ($u_{\rm{U}}$ and $u_{\rm{D}}$, respectively), density ($\r_{\rm{U}}$ and $\r_{\rm{D}}$, respectively), sound speed ($c_{\rm{U}}$ and $c_{\rm{D}}$, respectively) and local Mach number ($\Ma_{\rm{U}}$ and $\Ma_{\rm{D}}$, respectively). Hence, each truncated end is considered with an entirely separate model. Provided density (and the resulting sound speed) remain constant throughout each of the acoustic domains -- this excludes nonlinear acoustic waves and temperature variations at the boundary up- or downstream -- the acoustics travelling in these region may be determined by a representative time delay ($\t_{\rm{U}}$ and $\t_{\rm{D}}$) after the acoustic wave leaves the domain before it reenters with the same amplitude. Attenuation due to acoustics is neglected for these time scales by Stokes' law of sound attenuation~\cite{stokes1845TheoriesInternalFriction}. The time delay takes the form:
\begin{equation}
\t_{\rm{U},\rm{D}}
= \frac{L_{\rm{U},\rm{D}}}{c_{\rm{U},\rm{D}} - u_{\rm{U},\rm{D}}} + \frac{L_{\rm{U},\rm{D}}}{c_{\rm{U},\rm{D}} + u_{\rm{U},\rm{D}}}
= \frac{2 L_{\rm{U},\rm{D}}}{c_{\rm{U},\rm{D}}} \frac{1}{1 - \Ma_{\rm{U},\rm{D}}^2}.
\end{equation}
This is the time delay we will associate with the DNS in- and outflow. Thus, the tube length is represented by the delayed reentry of these acoustics into the inflow and outflow after a time $\t_{\rm{U}}$ and $\t_{\rm{D}}$, respectively. The acoustic period of the tube's fundamental mode is $t_{\rm{tube}} := 2(\t_{\rm{U}} + 2t_x + \t_{\rm{D}})$ for a closed-open tube with up- and downstream acoustic regions truncated with ADCBC in- and outflow. Provided the local Mach number remains low at either end of the tube, the associated delay $\t_{\rm{U},\rm{D}}$ can be approximated by $\t_{\rm{U},\rm{D}} \simeq 2 L_{\rm{U},\rm{D}} / c_{\rm{U},\rm{D}}$. This solves the linear acoustic problem in the tube up- and downstream of the flame to leading order in Mach number whenever the assumption of one-dimensional acoustics is valid. We refer to this as the \emph{Acoustic Delay Characteristic Boundary Condition} (ADCBC) method.

\subsection{Acoustic Delay Characteristic Boundary Conditions}

\begin{figure*}[t]
\centering
\includegraphics[scale=0.60]{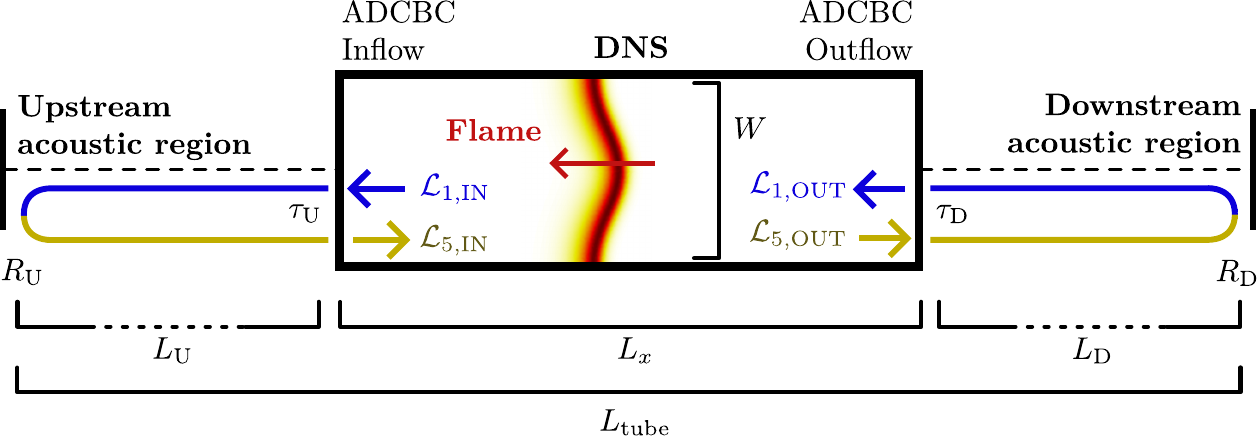}
\caption{Diagram of the ADCBC method.}
\label{fig:delay-model}
\end{figure*}

To employ the acoustic delay model within the NSCBC formulation, we assume for the moment we have access to the full history of outgoing acoustics, $\cl{L}_-(t, y)$ at a vertical boundary in two-dimensions. Continuing with this two-dimensional example, we can impose the delayed acoustic reentry:
\begin{equation}
\cl{L}_+(t, y)
= \cl{L}_+^{\rm{NR}}(t, y)
+ \cl{L}_+^{\rm{Delay}}(t, y)
\end{equation}
where
\begin{subequations}
\begin{align}
\cl{L}_{+, \rm{IN}}^{\rm{Delay}}(t, y)  &:= R_{\rm{U}} \cl{L}_{-, \rm{IN}}(t - \t_{\rm{U}}, y),\\
\cl{L}_{+, \rm{OUT}}^{\rm{Delay}}(t, y) &:= R_{\rm{D}} \cl{L}_{-, \rm{OUT}}(t - \t_{\rm{D}}, y).
\end{align}
\end{subequations}
The first terms $\cl{L}_+^{\rm{NR}}$ are the required condition described in \sect{sec:nonreflect} to stop the outgoing acoustic from being immediately reflected. For a perfectly reflecting acoustically closed upstream end $R_{\rm{U}} = 1$ and open downstream end $R_{\rm{D}} = -1$. Under this model, we assume implicitly that each location on the boundary has its own one-dimensional acoustic approximation. For the rest of this work we use a simpler formulation by averaging uniformly along the boundary:
\begin{subequations} \label{eqn:L-delay-form}
\begin{align}
\cl{L}_{+, \rm{IN}}^{\rm{Delay}}(t, y)  &:= R_{\rm{U}} \overline{\cl{L}_{-, \rm{IN}}}(t - \t_{\rm{U}}), \\
\cl{L}_{+, \rm{OUT}}^{\rm{Delay}}(t, y) &:= R_{\rm{D}} \overline{\cl{L}_{-, \rm{OUT}}}(t - \t_{\rm{D}}).
\end{align}
\end{subequations}
The operator $\overline{\phi}(t)$ represents averaging over in- or outflow boundary values of $\phi(t, y)$ respectively:
\begin{equation}
\overline{\phi}(t) = \frac{1}{W} \int_{-W/2}^{W/2} \phi(t, y) \dd{y}.
\end{equation}
This corresponds to a single one-dimensional approximation being made for each boundary, which is valid only for computational domains with width $W \ll L_{\rm{tube}}$. A diagram for this model is shown in \fig{fig:delay-model}.

When the acoustic field is strong enough for acoustic velocity to exceed inflow velocity, the up- and downstream boundaries each oscillate with the acoustics between in- and outflow. To account for this, we simply check the sign of $u$ and apply the LODI and diffusive conditions correspondingly. Switiching between conditions remains sufficiently smooth in time for time integration, provided $u$ remains sufficiently smooth in time.

\subsubsection{Discretisation}

In this manuscript, the SUNSET code is used to solve the governing equations~\cite{king2024MeshFreeFrameworkHighOrdera}. The code uses the Local Anisotropic Basis Function Method (LABFM)~\cite{king2020HighOrderDifference} to discretise the domain as an unstructured set of collocation points with node spacing $\d x$. LABFM is a high-order, mesh-free generalisation of centered finite differences. In this section, fourth-order consistent discretisations of first and second order derivative operators are used. Time integration is performed via the third-order Runge-Kutta of~\cite{kennedy2000LowStorageExplicitRunge} (coined RK3(2)4[2R+]C in their work). Adaptive time stepping is constrained in all cases by the CFL condition $\d t = \max\{ \d x / |\vb{u}| + c \} / 2$ due to the relatively large diffusive length scales in the modelled idealised flows. As with other high-order collocated methods, the solution must be de-aliased~\cite{orszag1971EliminationAliasingFiniteDifference}. This is done by applying a high-order hyper-viscosity operator to every field in the domain after each time step~\cite{king2022HighOrderSimulationsIsothermal}. Zero normal flux conditions are enforced on the relevant diffusive terms at in- and outflows to close the parabolic problem according to~\cite{sutherland2003ImprovedBoundaryConditions}.

When implementing into a Navier-Stokes solver, the ADCBC strategy employed approximates $\overline{\cl{L}}_-(t)$ by a piecewise constant function $\tilde{\cl{L}}(t)$:
\begin{equation} \label{eqn:disc-L}
\tilde{\cl{L}}_-(t) = \overline{\cl{L}}_-(t^l)
\end{equation}
for $t \in [t^l, t^{l+1}]$. Equation \equ{eqn:disc-L} is used in the solver in place of $\overline{\cl{L}}_-$ in $\equ{eqn:L-delay-form}$. Due to the dynamic time stepping describe above, we cannot guarantee a constant sampling period of $\overline{\cl{L}}_-$ values. We instead bound our sample times:
\begin{equation}
\d t_{\rm{sample}} < t^{l + 1} - t^l < \d t_{\rm{sample}} + \d t.
\end{equation}
Unless otherwise stated, a value of $\d t_{\rm{sample}} = 4 \d t_0$ is used, where $\d t_0$ is the initial time step. These values are stored in a memory buffer for $t^l$ and $\overline{\cl{L}}(t^l)$ values.


\subsection{Preliminary Tests}

We consider an inert fluid comprised of a single species, corresponding roughly to a stoichiometric methane-air mixture at atmospheric temperature and pressure. The same fluid properties are used in all examples:
\begin{equation}
M=28~\rm{kg}~\rm{kmol}^{-1},\quad
c_p=1.1~\rm{kJ}~\rm{kg}^{-1}~\rm{K}^{-1},\quad
\m=18~\text{{\textmu}g}~\rm{m}^{-1}~\rm{s}^{-1},\quad
\Pr=0.7
\quad\text{and}\quad
\Le = 1,
\end{equation}
where the non-dimensional Prandtl number $\Pr:=\m c_p / \l$ and Lewis number $\Le:=\l/(\r c_p D)$. Initial inflow conditions are:
\begin{equation}
u_{\rm{IN}}=0.2~\rm{m}~\rm{s}^{-1},
\quad
T_{\rm{IN}}=298~\rm{K}
\quad \text{and} \quad
p_{\rm{IN}}=1~\rm{bar}.
\end{equation}
This results in a uniform sound speed of $c\simeq348$~m~s$^{-1}$ in the domain. In this section the flow is initialised with uniform horizontal velocity with a superimposed acoustic perturbation. The discretisations are one-dimensional and equispaced with spacing $\d\/x = 30$~{\textmu}m (very fine for the relevant acoustic waves, but reasonable considering the diffusive and reactive length scales typical for combustion). Below we test the ADCBC approximation for in- and outflows. In the flames simulations in \sect{sec:5}, boundary averaging is still used. This is reasonable as long as the transverse acoustic eigenmodes of the domain are negligible.

\subsubsection{Reflected Travelling Acoustic Wave}

\begin{figure*}[t]
\begin{subfigure}{0.99\textwidth}
\centering
\includegraphics[scale=0.45]{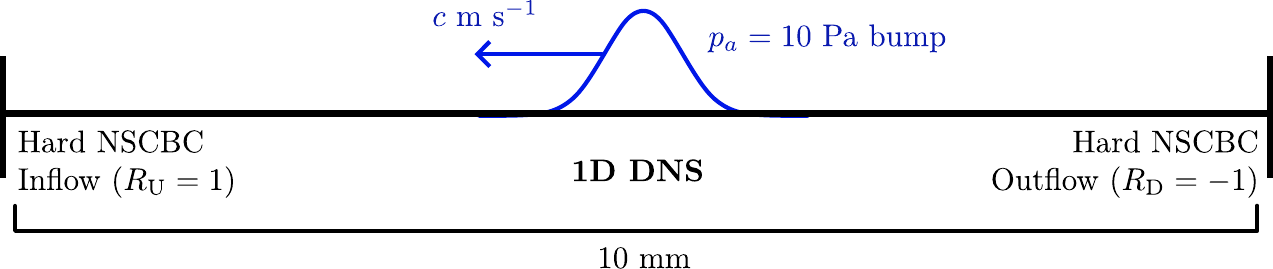}
\caption{}
\label{fig:adcbc-in-outa}
\end{subfigure}

\begin{subfigure}{0.99\textwidth}
\centering
\includegraphics[scale=0.45]{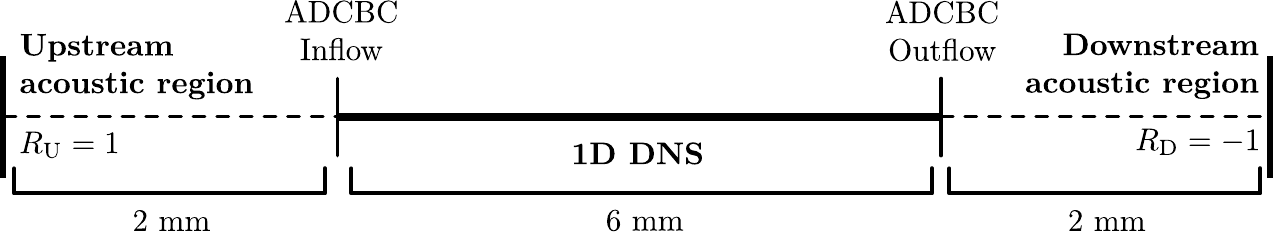}
\caption{}
\label{fig:adcbc-in-outb}
\end{subfigure}

\begin{subfigure}{0.99\textwidth}
\centering
\includegraphics[scale=0.40]{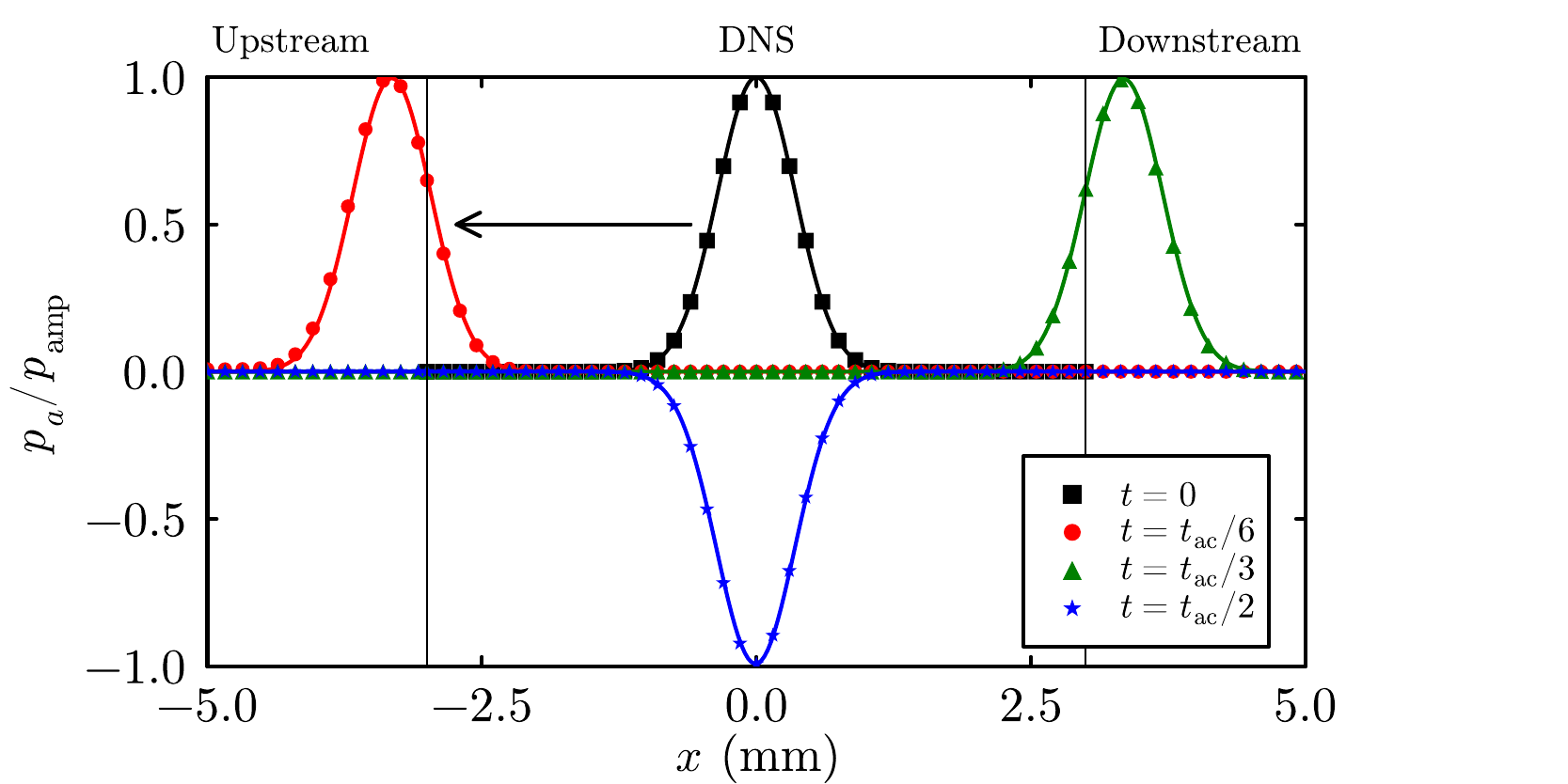}
\caption{}
\label{fig:adcbc-in-outc}
\end{subfigure}
\caption{Diagrams of the single reflection test case (a) without and (b) with ADCBC method in- and outflows, not to scale. (c) acoustic pressure amplitude over half an acoustic period. Symbols show results with ADCBC method in- and outflows and lines show results without (full DNS).}
\label{fig:adcbc-in-out}
\end{figure*}

We simulate a travelling acoustic wave which travels upstream from the centre of the DNS domain, reflecting off the up- and downstream boundary before returning to its original location. The initial acoustic disturbance is given by:
\begin{subequations}
\begin{align}
p_a(t = 0, x) &= p_{\rm{amp}} \exp\left[- \left( \frac{x - x_{\rm{centre}}}{x_{\rm{radius}}} \right)^2 \right], \\
u_a(t = 0, x) &= - \frac{1}{\r_0 c_0} p_a(t = 0, x),
\end{align}
\end{subequations}
with parameters $p_{\rm{amp}} = 10$~Pa, $x_{\rm{centre}} = 0$~mm and $x_{\rm{radius}} = 1$~mm. The DNS domain is $L_x = 6$~mm long, centred at the origin, with $L_{\rm{U}} = L_{\rm{D}} = 2$~mm, to model a total $L_{\rm{tube}} = 1$~cm of tube length where $R_{\rm{U}} = 1, R_{\rm{D}} = -1$. Results are compared to full DNS simulations of the whole domain $L_{\rm{tube}} = L_x = 1$~cm using perfectly reflecting inflows, $\cl{L}_+ = \cl{L}_-$ and perfectly reflecting outflows, $\cl{L}_+ = -\cl{L}_-$. In both cases the upstream boundary condition should enforce $u_{a, \rm{U}} \equiv 0$ and downstream boundary condition should enforce $p_{a, \rm{D}} \equiv 0$. This geometry is shown in \fig{fig:adcbc-in-outa} and \fig{fig:adcbc-in-outb}. When $t < 0$, $\tilde{\cl{L}}_-(t) = 0$ by assumption.

\fig{fig:adcbc-in-outc} shows results after half an acoustic period, or a single up- and downstream reflection. Both test cases with and without ADCBC method in- and outflows display accuracte reflection of the pressure disturbance. Using output in- and outflow ADCBC memory buffers, we are able to reconstruct the stored one-dimensional acoustic field from the time series data in the up- and downstream regions, respectively. This is shown by the markers in \fig{fig:adcbc-in-outc} in the up- and downstream regions.

\subsubsection{Convergence over Many Reflections} \label{sec:convmanyrefl}

\begin{figure*}[t]
\begin{subfigure}{0.99\textwidth}
\centering
\includegraphics[scale=0.45]{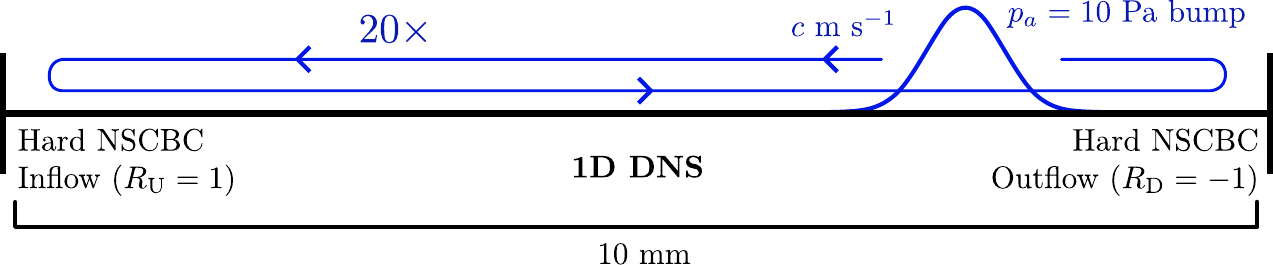}
\caption{}
\label{fig:in-bump-conv-testa}
\end{subfigure}

\begin{subfigure}{0.99\textwidth}
\centering
\includegraphics[scale=0.45]{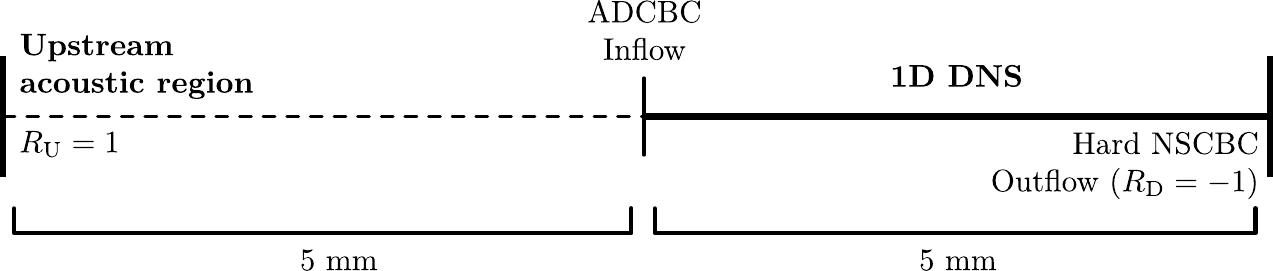}
\caption{}
\label{fig:in-bump-conv-testb}
\end{subfigure}

\begin{subfigure}{0.99\textwidth}
\centering
\includegraphics[scale=0.34]{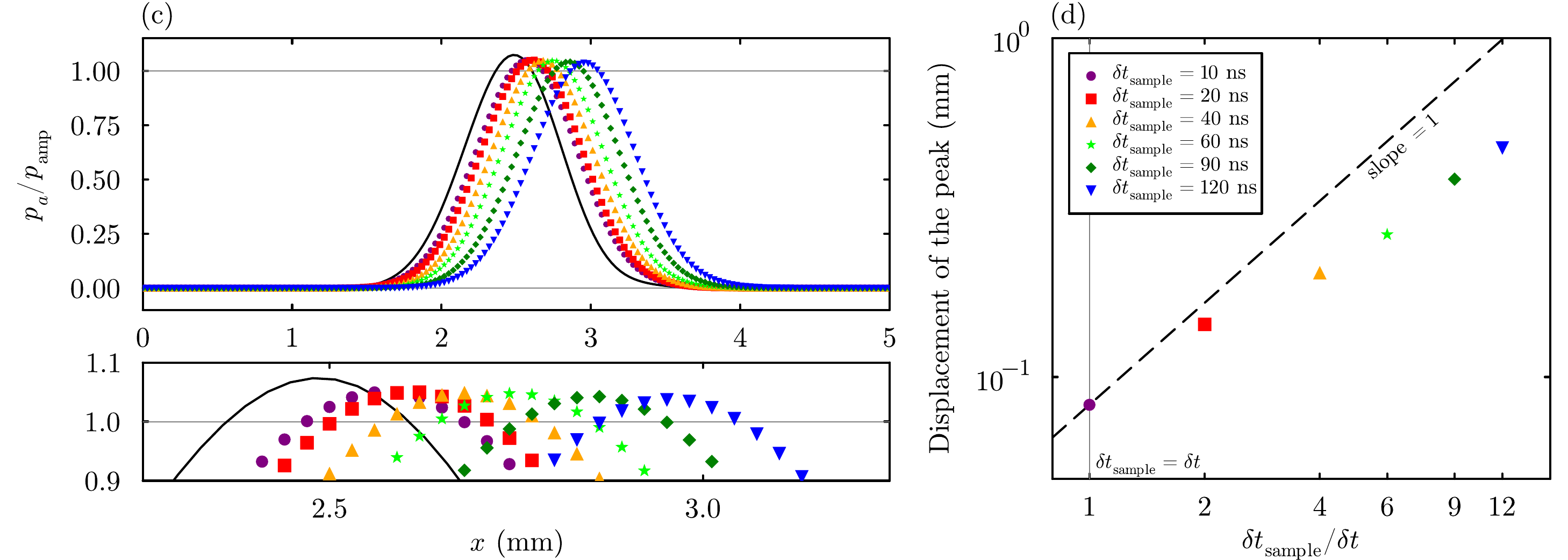}
\end{subfigure}
\caption{Diagrams of the many reflections test case (a) without and (b) with ADCBC inflow. (c) acoustic pressure amplitudes after 20 up- and downstream reflections for a range of ADCBC sample periods. The black line shows full DNS results and coloured markers show results using ADCBC inflow as sample time varies. (d) displacements of the peaks of ADCBC simulations away from the full DNS simulation, for each sample period.}
\label{fig:in-bump-conv-test}
\end{figure*}

With the same acoustic parameters as above, we test the response of the ADCBC method after many reflections. We truncate the upstream half of the domain with the ADCBC method such that $L_x = 5$ mm. The full DNS domain is the same as above. Ideally, ADCBC truncation will allow perfect reconstruction of the wave upon its reentry to the domain. However, due to the sampling of $\overline{\cl{L}}_-$ values and local truncation error from time integration at the boundaries, this will not be the case. As a result of discretisation error, we expect the discretised wave structure to deteriorate more rapidly in the full DNS case. Hence, the solutions with and without ADCBC truncation are expected to diverge in time, notwithstanding error introduced by the ADCBC method. We instead compare the position of acoustic disturbance peaks and require these to converge as ADCBC sample times decrease. Since the reconstructed waves are delayed due to imperfect time integration of $\cl{L}_-$ values, this imposes a displacement of the wave peak, introducing a finite error to the measured acoustic wave frequency.

\fig{fig:in-bump-conv-test}c and d show results after 20 upstream acoustic reflections, with displacements of the wave peak converging to less than $3\d x$ from the full DNS results. Note that the acoustic disturbance has retained its structure in all cases. Results indicate that, in the worst case, an approximate 0.05~mm peak displacement is imposed after each upstream reflection. This is noticeable over a few reflections in this short domain, but in a longer e.g.~10~cm domain remains negligible when compared to the acoustic wavelengths associated with a longer domain.

\subsubsection{Standing Acoustic Wave} \label{sec:standingwave}

\begin{figure*}[t]
\centering
\includegraphics[scale=0.34]{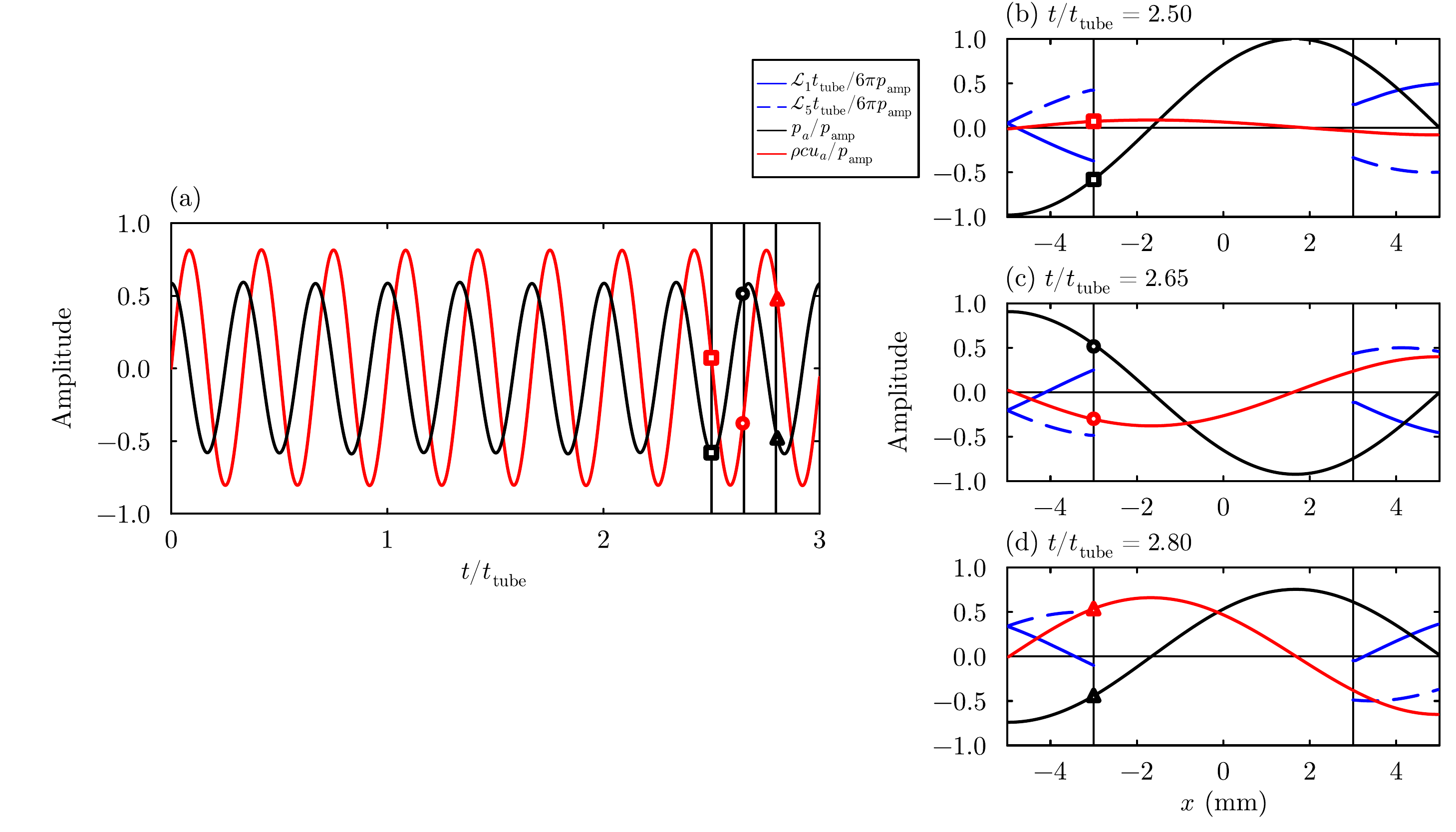}
\caption{(a) time series of acoustic inflow velocity and pressure. (b, c, d) acoustic fields in the whole domain at three snapshots over an acoustic period. Markers in (a) correspond to markers at the DNS inflow in (b, c, d).}
\label{fig:wave-later}
\end{figure*}


Excited acoustic modes in an extrinsic thermoacoustic instability exist as standing waves with a density discontinuity at the flame. Since the ADCBC method only presumes constant sound speed throughout the truncated region, the ADCBC method in a truncated hot and cold domain operate in fundamentally the same way. We test standing acoustic waves in the truncated domain described in \fig{fig:adcbc-in-outb}. Initial conditions are:
\begin{subequations}
\begin{align}
p_a(t = 0, x) &= p_{\rm{amp}} \cos\left[ 2 \p \left( \frac{x - x_{\rm{IN}} - L_{\rm{U}}}{L_{\rm{tube}}} \right)  \left( \frac{2N_\k - 1}{4} \right) \right], \\
u_a(t = 0, x) &= 0,
\end{align}
\end{subequations}
where $p_{\rm{amp}}$ is the acoustic amplitude as before. $N_\k$ determines the wavenumber of the standing wave: if $N_\k = 1$ we have the one-quarter mode, if $N_\k = 2$ we have the three-quarter mode etc. The corresponding dimensional acoustic wavenumber is $\k = 2 \p (2N_\k - 1) / 4 L_{\rm{tube}}$. In the acoustic regions, the wave can be initialised by calculating local $\cl{L}_{1, 5}(t = 0, x)$ values and entering these values into the initial ADCBC memory buffer for these regions. \fig{fig:wave-later} shows the simulation after a few acoustic periods for the $N_\k = 2$, $p_{\rm{amp}} = 10$~Pa case. Clearly $u_{a, \rm{U}} = 0$ and $p_{a, \rm{D}} = 0$ as desired. No noticeable change in amplitude is observed over the time simulated and no other acoustic modes are excited. Consistent with the results in the previous \sect{sec:convmanyrefl}, the correct standing wave frequency is reproduced.

\section{Drift Control} \label{sec:4}

\begin{figure*}[t]
\centering
\includegraphics[scale=0.34]{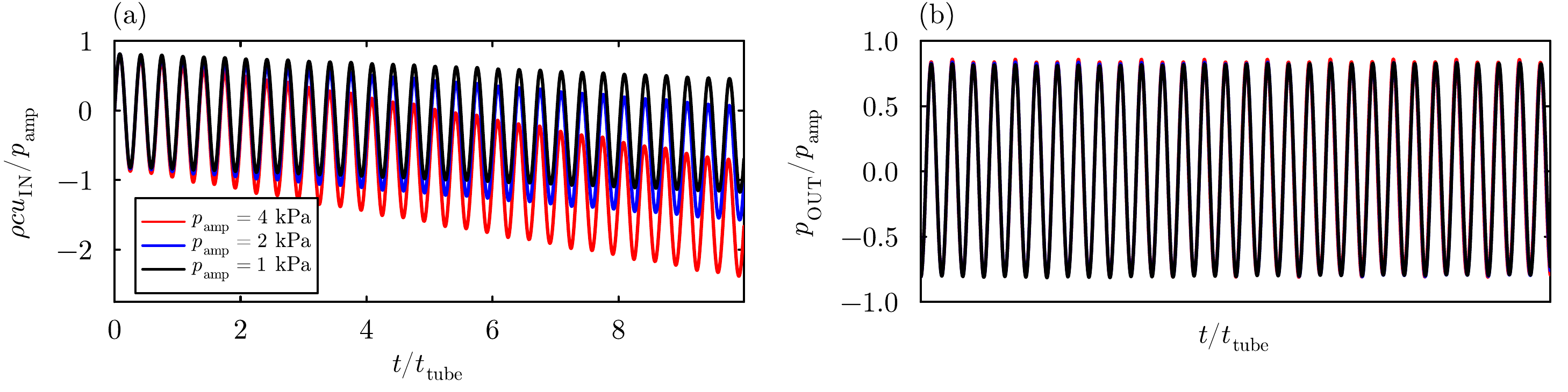}
\caption{Drifting acoustic (a) inflow velocity and (b) outflow pressure for three acoustic amplitudes, normalised by acoustic amplitude.}
\label{fig:inout-drift-test-scaling}
\end{figure*}

The non-conservative nature of the above formulation allows for a small drift in inflow velocity and outflow pressure when the ADCBC method is used. This drift accumulates over time, becoming significant for thermoacoustic simulations with dynamics which develop over many acoustic periods. We refer to the rate of drift of the mean inflow velocity and outflow pressure values as $\o_{\rm{drift}}$. To exemplify this, we model the standing wave test case in \sect{sec:standingwave} with $N_\k = 2$ and $p_{\rm{amp}} = 1, 2, 4$~kPa. Drift rate grows proportional to acoustic frequency, hence the choice of $N_\k$ is unimportant. Results in \fig{fig:inout-drift-test-scaling} show a nonlinear growth of the rate of growth $\o_{\rm{drift}}$ as $p_{\rm{amp}}$ increases. For every value of $p_{\rm{amp}}$ shown, the outflow pressure drift rate is minimal. Placing the truncated DNS boundary at a node of acoustic pressure or velocity results in negligible drift rate $\o_{\rm{drift}}$. This implies a proportionality in the drift rate with both $p_a$ and $u_a$:
\begin{equation}
\o_{\rm{drift}} \propto p_a u_a \propto p_{\rm{amp}}^2.
\end{equation}
This quadratic dependence is consistent with the results shown in \fig{fig:inout-drift-test-scaling}. In the following section, we develop a new model for linear relaxation terms to remove drift in an environment with persisting acoustics and a constant drift rate.

\subsection{A New Linear Relaxation Model}

\begin{figure*}[t]
\centering
\includegraphics[scale=0.38]{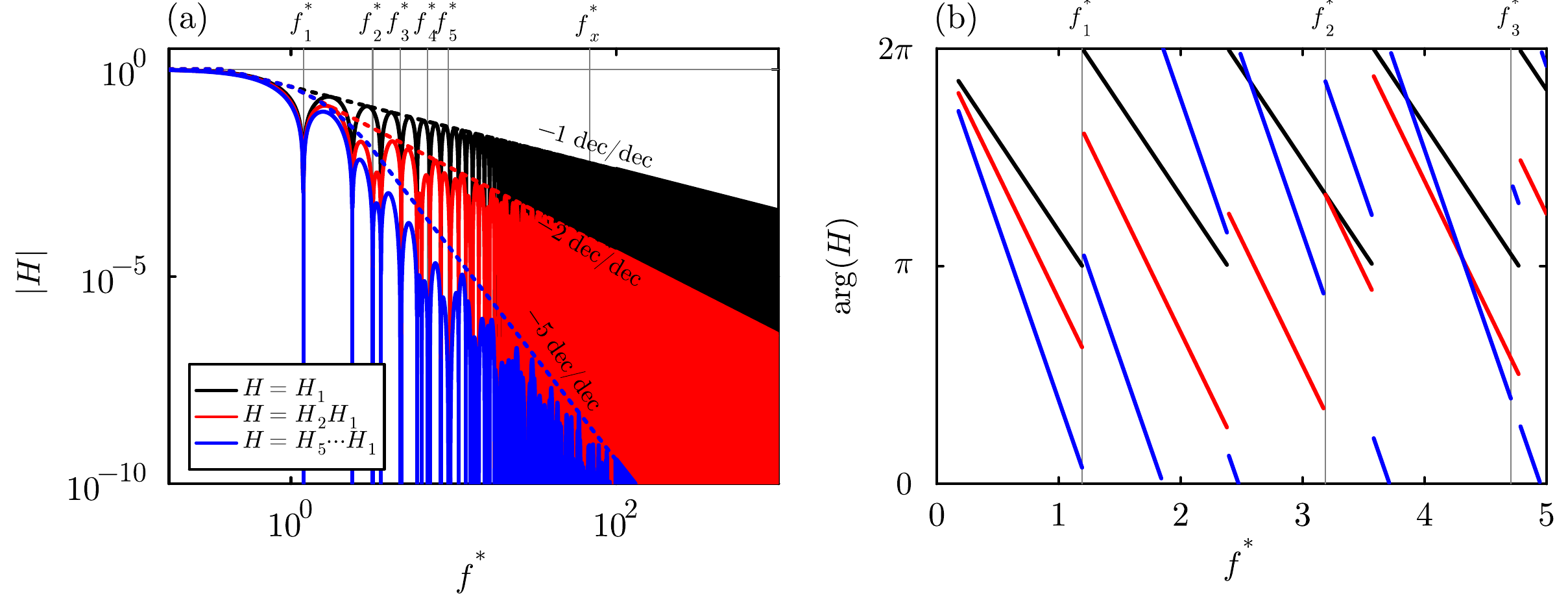}
\caption{The (a) magnitude and (b) argument of the moving averages transfer function for one (black), two (red) and five (blue) moving averages. The dotted lines are the envelopes $|\hat{H}_{\rm{env}}|$.}
\label{fig:transferH}
\end{figure*}

In the case of the truncated domains we model with the ADCBC method, acoustic domains are much longer than the DNS domain by design, $L_x \ll L_{\rm{tube}}$. This precludes using CLR to remove this drift as the fundamental mode and harmonics of the tube would undergo almost perfect reflection at the DNS boundaries. To account for this, we introduce the modified coefficient $\tilde{K} = 2 \s / t_{\rm{tube}}$. In doing so, we decrease the rate of convergence of the control response, but in proportion with the decreasing acoustic frequencies of the full domain. However, the oscillating acoustic signal at the boundaries remains difficult to effectively control. To treat this, we first perform a moving average of the in- or outflow signal over an averaging period $t_1$:
\begin{equation}
\langle s \rangle_1(t) := \frac{1}{t_1} \int_{t - t_1}^{t} s(t') \dd{t'}.
\end{equation}
It can be shown that the moving average of a signal $s(t)$ obeys the linear transfer function $\hat{H}_1 = \hat{H}_1(f)$ in frequency space such that $\hat{H}_1 \hat{s} = \hat{\langle s \rangle}$, where:
\begin{subequations}
\begin{equation}
\hat{H}_1(f) = \frac{1}{i2\p\!f t_1} \left( 1 - \exp(-i2\p\!f t_1) \right)
\end{equation}
has gain and phase:
\begin{align}
|\hat{H}_1|(f) &= \left| \frac{\sin(\p\!f t_1)}{\p\!f t_1} \right| \\
&\leq |\hat{H}_{\rm{env}}|(f) := \min \left(  1, ~ \frac{1}{\p\!f t_1} \right), \\
\arg(\hat{H}_1)(f) &= -\p \left(f t_1 - \lfloor f t_1 \rfloor\right) \Mod{2\p}.
\end{align}
\end{subequations}
So we ensure a gain of this transfer function which is less than one for frequencies $f > f^c := 1 / t_1$ whenever $\p > f_1^*$, where we denote non-dimensional frequencies via asterisks $f^* := f t_{\rm{tube}}$ and define the averaging frequency $f_1 = 1 / t_1$. Gain decreases in this region as $f_1^*$ decreases.

We now consider that this transfer function vanishes wherever $f t_1 = n$ for $n = 1, 2,$ and so on. This can be exploited by tuning the averaging period $t_1$ such that it equals the period of a chosen acoustic mode in the thermoacoustic system. In this way, the component of the signal, $s$, which contains that acoustic frequency is occluded when it is used in drift control. The integer $n$ represents the number of acoustic periods which fit into the averaging window. The averaging window $t_1$ corresponds to a moving average calculation which is delayed by $t_1 / 2$. Hence, the moving average acts as a delayed low-pass filter on the signal entering the relaxation terms. For one-dimensional acoustics without a density discontinuity, the eigenmodes are integer multiples of the fundamental mode -- for a closed-open tube these are the one-quarter mode, three-quarter mode and so on. In this case a single moving average with well chosen averaging period may be used to occlude all acoustic modes of the tube at once. Once a flame is introduced, however, the fundamental and harmonic acoustic modes are perturbed and become irrational multiples of each other. Hence, using a single moving average can only guarantee occluding a single acoustic mode in the signal $s$ at a time. To improve this, we can instead perform consecutive moving averages with averaging periods $t_1 > t_2 > \dots > t_N$. The resulting signal:
\begin{equation}
\cl{A}[s](t) := \langle\cdots\langle \langle s \rangle_1\rangle_2\cdots\rangle_N (t)
= \frac{1}{t_N \cdots t_2 t_1} \int^{t}_{t - t_N} \cdots \int^{t''}_{t'' - t_2} \int^{t'}_{t' - t_1} s(t') \dd{t'}\dd{t''}\cdots\dd{t}
\end{equation}
is the product of each moving average transfer function in frequency space:
\begin{equation}
\widehat{\cl{A}[s]\,}(f) = \left( \prod_{k = 1}^{N} \hat{H}_k(f) \right) \hat{s}(f) =: \hat{H}(f) \hat{s}(f),
\end{equation}
where the $\hat{H}_k$ is the transfer function corresponding to a moving average with period $t_k$ and $\hat{H}$ is the transfer function of $\cl{A}$.

These transfer functions are plotted over a range of frequencies in \fig{fig:transferH}a and b for one, two and five moving averages. The values of $t_k$ used are taken as the first five acoustic eigenmodes of a 100 cm closed-open tube containing a flame centred in the domain with density ratio over the flame $\r_{\rm{U}} / \r_{\rm{D}} = 7$. This corresponds to the first test case explored in \sect{sec:5}. \fig{fig:transferH}a illustrates the increasing attennuation of the averaged signal with frequency as more moving averages are used. These slopes are referred to as the damping rate, which increases by one for each moving average used for frequencies above the highest averaging frequency $f^*_k = t_{\rm{tube}} / t_k$ (we similarly define $f^*_x := t_{\rm{tube}}/t_x$). For a single moving average, we observe $|\hat{H}| \lesssim 10^{-0.5}$ for $f^* > f^*_1$ and that $|\hat{H}| \lesssim 10^{-2}$ for $f^* > f^*_x$. Performing more moving averages improves on this since $|\hat{H}| \lesssim 10^{-1.0}$ for $f^* > f^*_1$ and $|\hat{H}| \lesssim 10^{-8.5}$ for $f^* > f^*_x$. \fig{fig:transferH}b illustrates the piecewise linear relationship between the frequency and phase. Each jump discontinuity in the phase of $\hat{H}$ corresponds to a zero of its magnitude. Only a small section of the phase is shown for clarity.

At this point, if we use CLR on the averaged signal, we have a proportional (P) closed-loop feedback control term. As a result, for a signal subject to a constant drift rate of $\o_{\rm{drift}}$, we expect the proportional controller to stabilise toward a non-zero equilibrium error,
\begin{equation}
s_\infty := \frac{1}{2} \lim_{a\to\infty} \left(\max_{t>a}(\cl{A}[s](t)) + \min_{t>a}(\cl{A}[s](t))\right) \neq 0.
\end{equation}
This is exacerbated by the quadratic scaling of drift rate shown above, making it difficult to assign $\s$ \emph{a priori} such that $s_\infty$ remains small when the peak pressure amplitude of the thermoacoustic flame is difficult to predict. Hence, we introduce an integral (I) term and gain parameters $\h_P > 0$ and $\h_I > 0$:
\begin{equation} \label{eqn:PIRelax}
\cl{L}_+^{\rm{Relax}}(t, y)
:= \tilde{K} \bigg( \eta_{\rm{P}} \cl{A}[s] (t)
+ \frac{\eta_{\rm{I}}}{t_{\rm{tube}}}  \int_{0}^{t} \cl{A}[s] (t') \dd{t'} \bigg)
\end{equation}
for both in- and outflow relaxation terms (although the integral term for outflows is ignored later). We refer to this formulation as Averaged Proportional and Integral Linear Relaxation (APILR). In this way, we implement a PI controller on a low-pass filtered signal to numerically suppress the oscillating component. The resulting formulation including ADCBC terms is:
\begin{equation} \label{eqn:apilr_form}
  \cl{L}_+(t, y)
= \cl{L}_+^{\rm{NR}}(t, y)
+ \cl{L}_+^{\rm{Delay}}(t, y)
+ \cl{L}_+^{\rm{Relax}}(t, y)
\end{equation}

\begin{figure*}[t]
\centering
\includegraphics[scale=0.38]{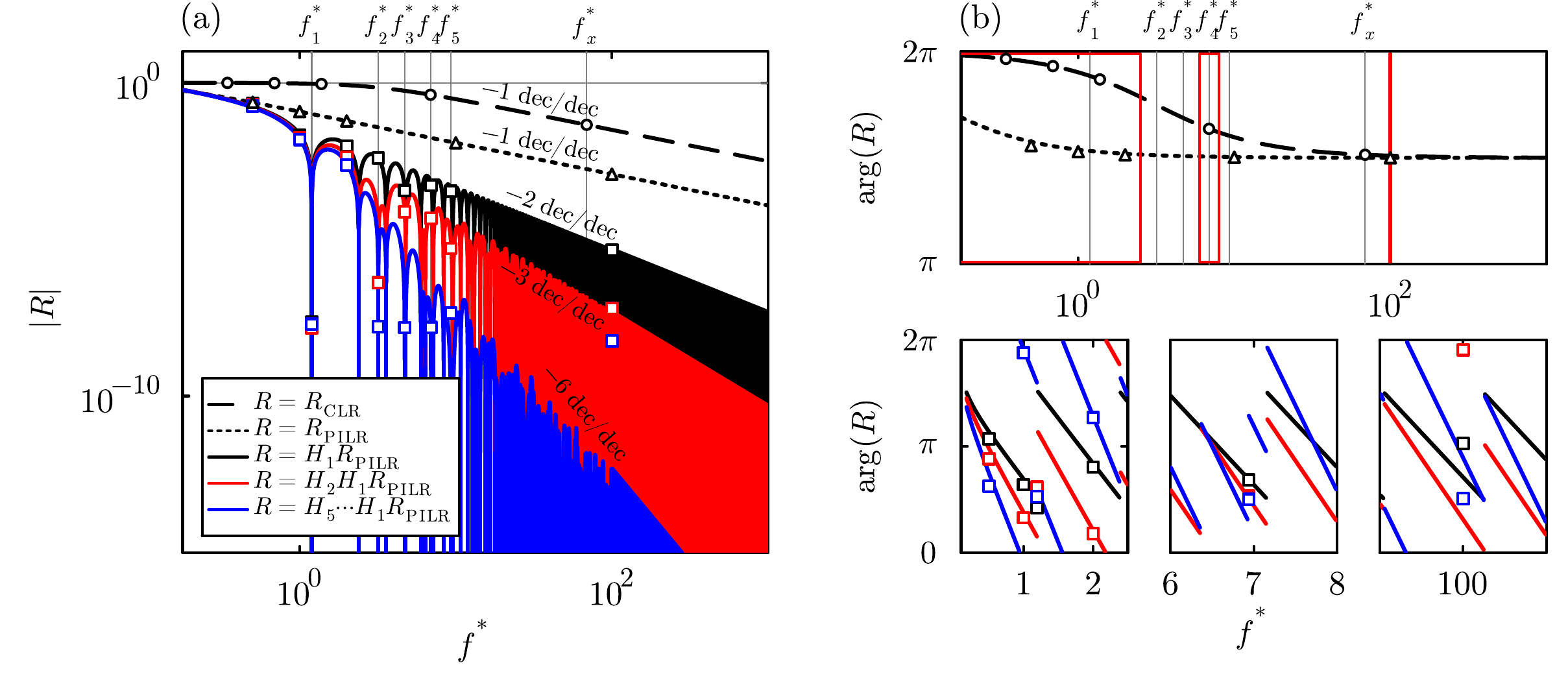}
\caption{The (a) magnitude and (b) argument of reflection coefficients $\hat{R}$ using three methods, CLR (dashed), PILR (dotted) and APILR (solid). APILR results are shown for one (black), two (red) and five (blue) moving averages. In all cases $\h_P = 2.65$ and $\h_I = 0.5$ are used. Scatter points show numerical data (black circles for CLR, black triangles for PILR, and black, red and blue squares for APILR using one, two and five moving averages, respectively).}
\label{fig:transferR}
\end{figure*}

When moving averages are not used, we instead use the name Proportional and Integral Linear Relaxation (PILR). Performing analysis similar to~\cite{selle2004ActualImpedanceNonreflecting}, one can show that the inclusion of an integral term to \equ{eqn:PIRelax} without moving averages results in the analytical reflection coefficient:
\begin{equation}
\hat{R}_{\rm{PILR}}(f) = \pm \frac{\h_I + (i \h_P) (2\p\!f t_{\rm{tube}})}{(\h_I - (2\p\!f t_{\rm{tube}})^2 / \s) + (i \h_P) (2\p\!f t_{\rm{tube}})}.
\end{equation}
where inflows have $+$ and outflows have $-$. It can be shown for all values $\h_I$ that $|\hat{R}_{\rm{PILR}}(f)| > 1$ whenever:
\begin{equation}
0 < f t_{\rm{tube}} < \frac{\sqrt{2 \s \h_I}}{2\p},
\end{equation}
which results in spurious acoustic energy production at the boundary. This is supressed by the introduction of the moving average filter, such that:
\begin{equation}
\hat{R}_{\rm{APILR}} \hat{s} = \hat{R}_{\rm{PILR}}\widehat{\cl{A}[s]} = (\hat{R}_{\rm{PILR}} \hat{H}) \hat{s}.
\end{equation}
For a closed-open tube we choose gains $\h_{P, \rm{IN}} = 2.65$, $\h_{I, \rm{IN}} = 0.5$, $\h_{P, \rm{OUT}} = 1$ and $\h_{I, \rm{OUT}} = 0$ by trial-and-error to achieve convergence to the target values in a few acoustic periods. These are the values used in all cases in this work. The value $\h_{I, \rm{OUT}} = 0$ is used as outflow pressure drifts negligibly from its target value over the physically relevant time scales. Higher gains result in faster response times, but introduce the risk of positive feedback due to the delayed averaging signal. This risk enforces a maximum value for these gains corresponding to the averaging periods. These scale with the acoustic tube time scale, $t_{\rm{tube}}$, so are not a concern for our test cases.

Reflection coefficients for CLR, PILR and APILR are shown in \fig{fig:transferR}. The inclusion of $R_{\rm{PILR}}$ in the reflection coefficients mean their damping rates have each increased by one. Notice also how the phase behaviour from $H$ carries over to $R_{\rm{APILR}}$, resulting in vanishing reflection coefficient under the APILR method at the chosen frequencies. Numerical results are obtained using a small, one-dimensional simulation with outflows injected with acoustics at a constant frequency $f$ as the tube length $t_{\rm{tube}}$ changes. The full in- or outflow conditions with injected acoustics is:
\begin{equation} \label{eqn:L-with-inj}
  \cl{L}_+
= \cl{L}_+^{\rm{NR}}
+ \cl{L}_+^{\rm{Delay}}
+ \cl{L}_+^{\rm{Relax}}
+ \cl{L}_+^{\rm{Inject}}
\end{equation}
where the non-injection terms are the same as in \equ{eqn:apilr_form} and the injected acoustic field is defined using \equ{eqn:single_char_prob}:
\begin{equation}
\cl{L}_+^{\rm{Inject}} = -\dv{t} J_+^{\rm{Inject}}
\end{equation}
Injected acoustics have the form $J_+^{\rm{Inject}}(t) := p_{\rm{amp}} \sin(2\p f t)$. Notwithstanding high $p_{\rm{amp}}$ values resulting in acoustic nonlinearity, the value of $p_{\rm{amp}}$ does not affect results due to linearity of the boundary formulation. For this verification of inflow relaxation terms, we use only the first and third terms of \equ{eqn:L-with-inj} for the inflow boundary and first and fourth term for the outflow boundary. Inflow reflection coefficients are then calculated directly as $\hat{R}(f) = \hat{\cl{L}}_+(f) / \hat{\cl{L}}_-(f)$. This corresponds to the method used by~\cite{selle2004ActualImpedanceNonreflecting} for outflow CLR. To implement each moving average, a memory buffer is used to store many uniformly sampled values of $s(t')$ for $t' \in [t - t_1, t]$. From \fig{fig:transferR}a we observe that introducing more moving averages reduces the gain by multiple orders of magnitude for signals at the averaging frequency. Numerical results to plateau at $|\hat{R}_{\rm{APILR}}| \simeq 10^{-8}$ due to errors introduced at the postprocessing stage.


\subsection{Non-Drifting Acoustic Waves}

\begin{figure*}[t]
\centering
\includegraphics[scale=0.38]{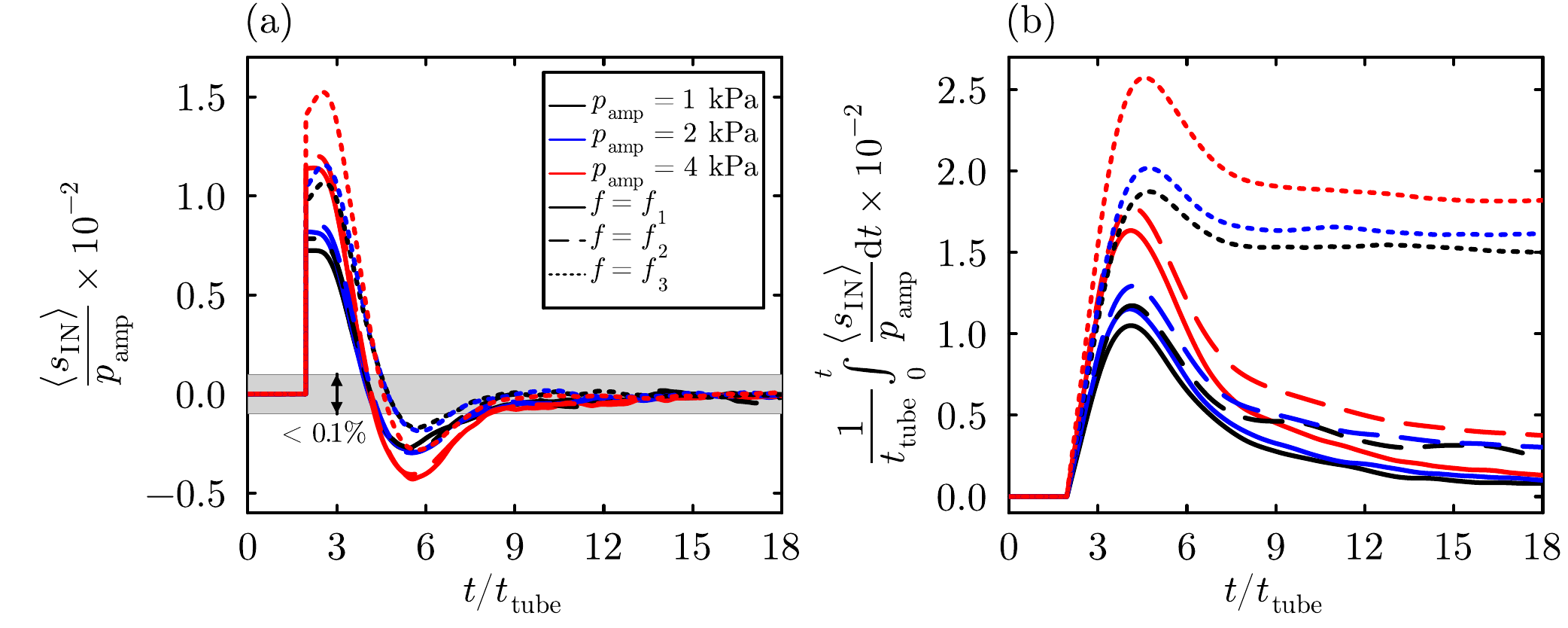}
\caption{Normalised (a) drift and (a) integral of drift for acoustic inflow velocity. Solid, dashed and dotted lines show response to incident acoustic waves at frequencies $f_1$, $f_2$ and $f_5$, respectively. Black, blue and red lines show response to incident acoustic waves of amplitude 1, 2 and 4 kPa, respectively.}
\label{fig:inout-drift-test}
\end{figure*}

\fig{fig:inout-drift-test} shows results when the APILR formulation is used for non-reflecting in- and outflows without ADCBCs and where the five moving averages are taken consecutively, as above. Results are similar when ADCBCs are employed with standing waves, except that reflected waves from the inflow leave the domain via the outflow. Target inflow velocity is $u^\rm{t} = 0.2$~m~s$^{-1}$. In all simulations, acoustics are filtered out by the APILR method (this is only possible when moving averages discretise the acoustic period into an integer number of samples). At all times, inflow drift values remain below 2\% the magnitude of the acoustic waves. After only $t > 8 t_{\rm{tube}}$, drift remains below 0.1\% for the remainder of the simulation. As a result, although integral values for $f = f_5$ differ from the lower frequency tests, all integral values in \fig{fig:inout-drift-test}b remain bounded. When the integral term is  excluded from the formulation, these integral values diverge as no attempt is made to control them. Similar results may be obtained for outflows which are not shown here as the outflows do not significantly drift.

\section{Simulation of Thermoacoustically Unstable Counterflow Flames} \label{sec:5}
In this section, we perform DNS of idealised premixed flames in closed-open tubes using ADCBC inflows and outflows. This tests the use of ADCBC truncation and APILR to reduce simulation cost of a thermoacoustically unstable flame. In the previous sections, the ADCBC method has been validated under one-dimensional inert conditions. As the ADCBC truncation only assumes constant fluid properties in the acoustic domain, this validation also applies to a one-dimensional flame simulation with cold inflow and hot outflow, provided the flame does not intersect the boundary; flashback and flamelets advecting through the outflow are prohibited under this model. Below we demonstrate usage of ADCBC truncation and APILR control terms in the context of thermoacoustically unstable flames.

To model the flame, we use the fluid properties and inflow conditions from \sect{sec:3} and impose a single irrerversible reaction step $\rm{R}\to\rm{P}$ from cold reactants R to hot products P. The fluid is determined by the same parameters used in the previous sections. We characterise this flame as an idealised representation of a stoichiometric methane-air flame at atmospheric temperature and pressure. The following flame parameters are used: heat release parameter $q := T_{\rm{adb}} / T_{\rm{IN}} - 1 = 6$ where $T_{\rm{adb}}$ is the adiabatic flame temperature, laminar flame speed $S_L$ and Zel'dovich number $\Ze := E_{\rm{act}} (T_{\rm{adb}} - T_{\rm{IN}}) / R_0 T_{\rm{adb}}^2$ where $E_{\rm{act}}$ is the dimensional activation energy. In the following section the values are used:
\begin{equation}
q = 6,
\quad
T_{\rm{adb}} = 2086~\rm{K}
\quad
S_L = 0.2~\rm{m}~\rm{s}^{-1}
\quad \text{and} \quad
\Ze = 5.
\end{equation}
The resulting sound speed in the hot products is $c_{\rm{D}} = c_{\rm{U}}\sqrt{1 + q} \simeq 778$~m~s$^{-1}$ and the laminar flame thickness is $L_f = 0.216$~mm.

In all cases in this section, we model a free flame centred in the acoustic domain. The DNS domain is $L_x = 2$~cm long and $W = 4$~mm wide with symmetry boundaries at the top and bottom of the domain and in- and outflow boundaries to the left- and right-hand side of the domain, respectively. The premixed reactant R enters from an acoustically closed upstream end, reacts in the DNS domain and product P leaves through an acoustically open downstream end. The two-dimensional DNS domain is discretised by approximately 50,000 degrees of freedom with discretisation length scale $\d x = 27$~{\textmu}m (eight points discretise one laminar flame thickness). In this section, eighth-order consistent LABFM discretisations are used. The APILR parameters are those chosen in \sect{sec:4}, where $\h_{\rm{P}} = 2.65$, $\h_{\rm{I}} = 0.5$ and we use five moving averages with averaging periods equal to the fundamental acoustic mode and first four harmonics.

\subsection{Primary Instability}

\begin{figure*}[t]
\centering
\includegraphics[scale=0.60]{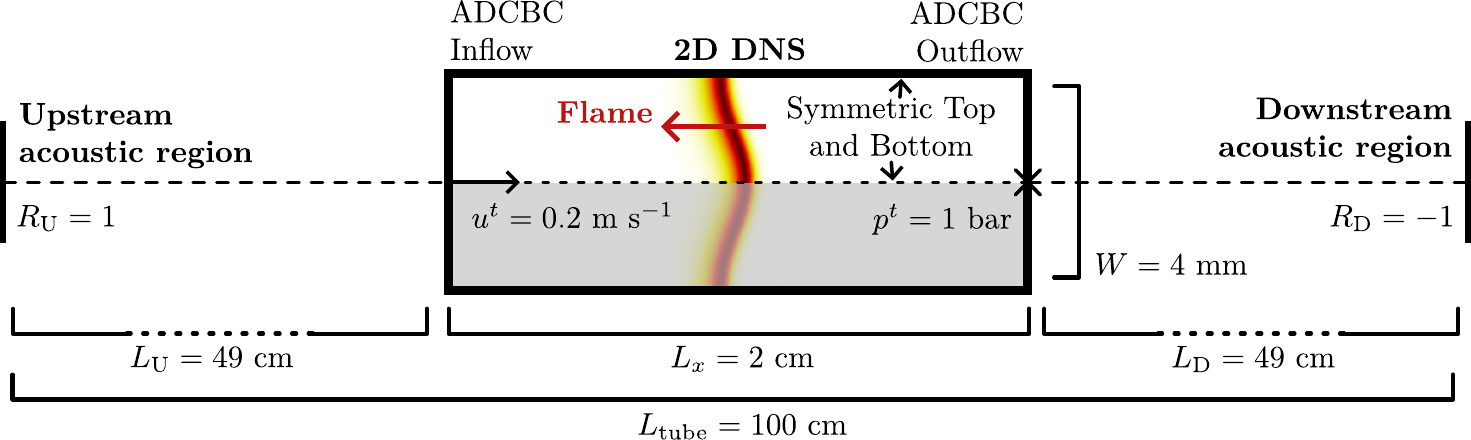}
\caption{Diagram of counterflow flame DNS and acoustic geometry. Not to scale.}
\label{fig:ff-domain}
\end{figure*}

First, we model the progression of primary thermoacoustic instability in a $L_{\rm{tube}} = 100$~cm long tube. Hence, $L_{\rm{U}} = L_{\rm{D}} = 49$~cm. A diagram of this geometry is shown in \fig{fig:ff-domain}. For a flame with $q = 6$ at the centre of the tube, we determine their one-dimensional, non-dissipative acoustic eigenmodes from analysis in \appx{ap:eigenmodes} to leading order in Mach number~\cite{clavin1990OnedimensionalVibratoryInstability}. The fundamental and first harmonic of the tube are the perturbed one- and three-quarter modes with frequencies $f_{1/4} \simeq 150$~Hz and $f_{3/4} \simeq 400$~Hz. This analysis only accounts for the modes of a stationary flame, which is not the case when the flame is curved and exceeds the inflow speed of 0.2~m~s$^{-1}$. The flame under primary thermoacoustic instability moves negligibly in the full acoustic domain. In cases where the flame moves a significant length along the tube, the acoustic modes in the tube instead increase in frequency with the moving flame. This is the case for many thermoacoustically unstable flame experiments, e.g. the downward propagating flames in~\cite{flores-montoya2022NonadiabaticModulationPremixedflame,gaton-perez2025MitigationThermoacousticInstabilities,gaton-perez2025WalldominatedTransitionPremixedflame}. The simulations were carried out over approximately four days in each case by decomposing the domain onto 48 cores of an AMD Genoa CPU (taking approximately 4,608 core hours total). Using ADCBC truncation, we have saved 98\% on the computational cost by truncating the upstream and downstream ends of the domain.

\begin{figure*}[t]
\centering
\includegraphics[scale=0.34]{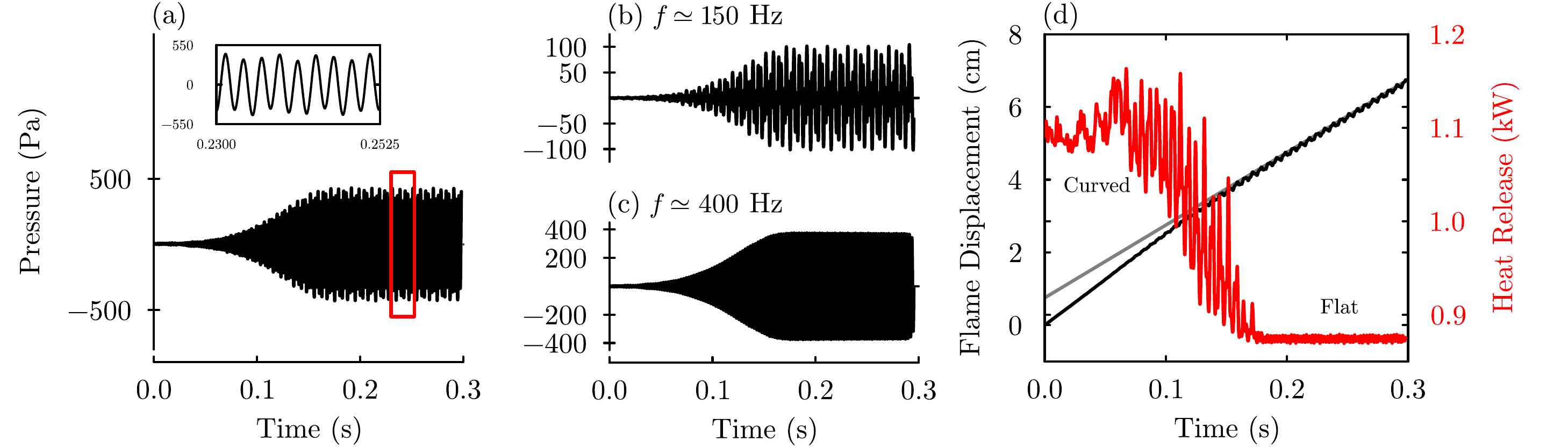}
\caption{(a) time series of inflow acoustic pressure and time series the (b) fundamental mode and (c) first harmonic. (d) time series of flame displacement (calculated as the total leftward flame motion from the upstream moving reactant reference frame) and integrated heat release rate.}
\label{fig:ff-stats}
\end{figure*}

\begin{figure*}[t]
\centering
\includegraphics[width=0.8\textwidth]{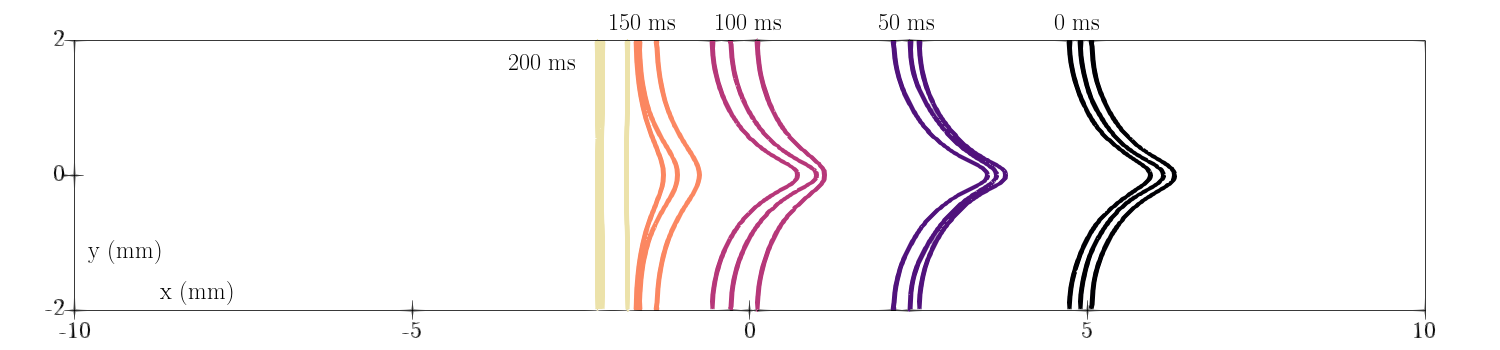}
\caption{Stroboscope of the flame in the DNS domain. Density isocontours are evaluated at $\r = 0.6$~kg~m$^{-3}$ over one fundamental acoustic period at phases $0$, $\p$ and $2\p$ at each of the times shown.}
\label{fig:ff-DL-flat-pv}
\end{figure*}

\fig{fig:ff-stats} shows time series of inflow acoustic pressure in both tubes as the thermoacoustic instability develops. The first harmonic dominates the acoustic signal at all times. All higher harmonics of the flame tube are negligible. The system experiences linear instability growth whilst the flame remains curved. As the flame flattens, nonlinear effects saturate the primary instability growth, resulting in a plateau of acoustic pressure for both the fundamental mode and first harmonic. This results in quasiperiodic limit cycle oscillations and no secondary instability growth is observed. These results corroborate those observed repeatedly in the literature where a stable plateau amplitude is reached (e.g.~\cite{searby1992AcousticInstabilityPremixed} among others). As the flame flattens, integrated heat release recedes to the heat release of a steady unstretched flat flame in a non-acoustic environment. The flame motion and flattening is shown in \fig{fig:ff-DL-flat-pv}. As the acoustic amplitude increases, the flame shape deforms over acoustic periods in response to the acoustic field. As the mean inflow is maintained due to APILR drift control, the curved flame advects toward the inflow boundary steadily until the acoustic plateau is reached and the flame flattens. Subsequently, the counterflow and flame flattening means the flame position in the DNS domain moves only due to acoustic vibration.

\begin{figure*}[t]
\centering
\includegraphics[scale=0.34]{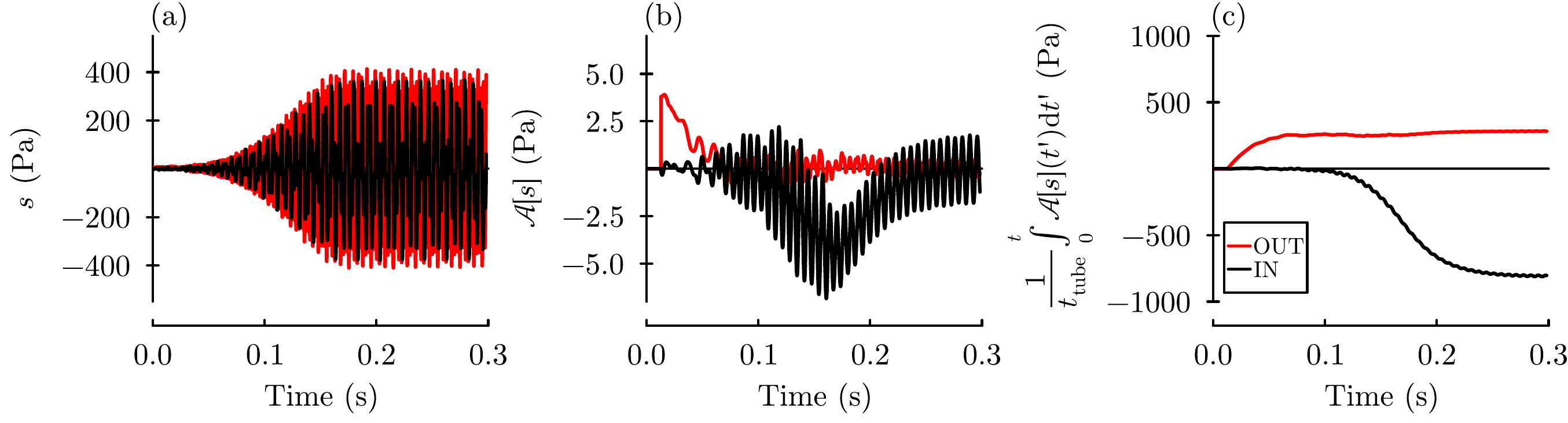}
\caption{(a) time series (b) drift and (c) drift integral of acoustic inflow velocity (black) and outflow pressure (red).}
\label{fig:ff-drift}
\end{figure*}

\fig{fig:ff-drift} shows time series of DNS inflow velocities and outflow pressures, as well as the moving average of drift in these variables away from their target values and integrals thereof. Outflow pressure data are not of major concern as drift 1 Pa oscillations are negligible in a background pressure of 1~bar. Inflow velocities, however, are more delicate since even small drift of mean inflow velocity from 0.20 to 0.21~m~s$^{-1}$ (corresponding to values $\cl{A}[s] \simeq 4$~Pa) results in rightwards advection of the flame of 1 cm after 1~s. This is a large enough displacement to move a flame centred in the 2~cm long DNS domain to the inflow boundary. Though, since the inflow drift returns to $\cl{A}[s] \simeq 0$~Pa after the plateau is reached, this is not a concern. For future simulations involving a dynamical target inflow $u^\rm{t} = u^\rm{t}(t)$, this may become a concern as we attempt to contain parametrically unstable flames for longer. We also observe acoustic oscillation in the drift signal, which is surprising as the moving averages are intended to occlude this in the drift control signal. This is likely due to a difference between chosen frequencies $f_k$ calculated from the acoustic eigenmodes \appx{ap:eigenmodes} and the true thermoacoustic frequencies in the simulation. This could be avoided by instead estimating the dominant frequencies on the fly. A consequence of this is the APILR boundaries introducing artificial reflections, although these do not spoil the results due to the short simulation time scale relative to the acoustic periods. For longer domains this will remains true as fewer acoustic oscillations result in fewer artificial reflections whilst more of the domain is truncated due to the ADCBC scheme. These results also demonstrate the boundary conditions remain stable as normal boundary velocity changes sign -- that is, when inflows become outflows and vice versa.

\subsection{Secondary Instability}

Next, we discuss the scarcely simulated cellular secondary thermoacoustic instability. Results from previous numerical works exhibiting this phenomena use more complex chemistry and transport~\cite{jun2023ParametricInstabilityPropagating} or include wall boundaries~\cite{gonzalez1996AcousticInstabilityPremixed,rodriguez-gutierrez2026EffectInducedOuter,chen2026AcousticResponseAsymmetric}. In this instance, we investigate the response of a premixed flame to acoustic forcing, in the style of~\cite{searby1991ParametricAcousticInstability} where we instead employ idealised physics to better understand the phenomena. This means using the ADCBC and APILR schemes with acoustic injection at the outflow as in~\equ{eqn:L-with-inj}. In this way, we measure the response of the premixed flame to acoustic waves at a chosen amplitude.

First, we model the response of a flat flame to harmonic transverse perturbations. Then, for validation we compare the qualitative behaviour of the DNS parametric instability response to results from the asymptotic theories. Consider the non-dimensionalisation of space, velocity, time, density and acoustic pressure by characteristic values $W / 2\p$, $S_L$, $W / 2\p S_L$, $\r_{\rm{U}}$ and $\r_{\rm{U}} S_L^2$, respectively. The behaviour of wavenumber $k$ disturbances to a flat flame are described by the damped Mathieu equation~\cite{assier2014LinearWeaklyNonlinear,searby1991ParametricAcousticInstability}:
\begin{equation} \label{eqn:AW-mathieu}
A \pdv[2]{\ftvar{F}}{t} + B(k) \pdv{\ftvar{F}}{t} + [C_1(k) \cos(2\p\!f t) + C_0(k)] \ftvar{F} = 0
\end{equation}
where $F = F(t, y)$ is the equation describing the flame front and $\ftvar{F} = \ftvar{F}(t, k)$ is its Fourier transform in $y$. A value $k = 1$ represents a perturbation of wavelength $W$ to the flame front, a value $k = 2$ represents a perturbation of wavelength $W / 2$ and so on. We define the coefficients:
\begin{equation}
A      = 1 + \frac{1}{1 + q},\quad
B(k)   = 2|k| + \frac{A q}{\g} k^2, \quad
C_1(k) = \frac{p_{\rm{amp}}}{2\p\!f \Ma_{\rm{U}}} |k|
\quad \text{and} \quad
C_0(k) = -2qk^2 + \frac{2 q}{\g} |k|^3.
\end{equation}
The non-dimensional tube width, $\g$ is defined as:
\begin{equation}
\g(W) := \frac{q W}{2\p  L_{\rm{th}} \Mk},
\quad \text{where} \quad
L_{\rm{th}} := \frac{\m }{\r S_L \Pr}
\end{equation}
is the thermal length scale and $\Mk$ the Markstein number. Here, we assume an existing acoustic field of amplitude $p_{\rm{amp}}$ and an incident acoustic wave of frequency $f$, modelled by the cosine forcing term in \equ{eqn:AW-mathieu}. As described in~\cite{searby1991ParametricAcousticInstability}, this is a reasonable approximation for a flame under primary thermoacoustic instability.

It is a well known result that solutions to \equ{eqn:AW-mathieu} assuming zero acoustic amplitude and the Michelson-Sivashinsky (MS) equation~\cite{sivashinsky1977NonlinearAnalysisHydrodynamic,michelson1977NonlinearAnalysisHydrodynamic}, which models the DL flame wrinkling effect, are both stable for non-dimensional widths of $\g < 2$,~\cite{vaynblat2000StabilityPoleSolutionsa,vaynblat2000StabilityPoleSolutions}. Evaluating Markstein numbers using the formula of~\cite{clavin1982EffectsMolecularDiffusion}, however, yields a marginally unstable flame with critical value $\g^c := \g(W^c) = 2$ at critical width $W^c \simeq 0.24$ mm; whereas the true stability limit, measured from DNS trial-and-error, was found to occur at $1.5~\rm{mm} < W^c < 2.0~\rm{mm}$. This is corroborated by previous literature suggesting that calculated $\g$ values may not be as important for quantitative predictions~\cite{assier2014LinearWeaklyNonlinear,assier2014CombustionInstabilityModel}. To instead choose $\g$ based off qualitative observations, we ensure the theoretical planar stability at the critical value $\g^c = 2$ agrees with the stability limit $W^c$ found by numerical trial-and-error, such that $\g(W) = 2W / W^c$. The predictions made below do not change as $W^c$ varies between 1.5~mm and 2.0~mm.

\begin{figure*}[t]
\centering
\includegraphics[scale=0.34]{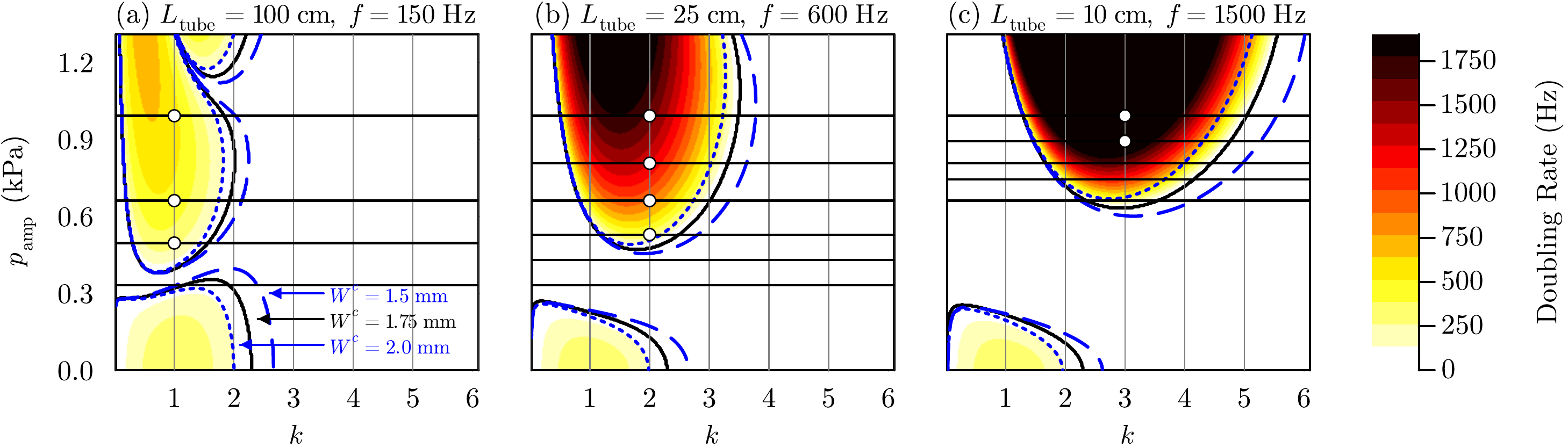}
\caption{Instability regions for the damped Mathieu equation, \equ{eqn:AW-mathieu}, in parameter space $(k, p_{\rm{amp}})$. The frequency $f$ is chosen as the fundamental mode in a (a) 100~cm long domain, (b) 25~cm long domain and a (c) 10~cm long domain. Marginal stability contours are shown for $W^c = 1.5$~mm (blue dashed line), $W^c = 1.75$~mm (black solid) and $W^c = 2.0$~mm (blue dotted). Instability regions for $W^c = 1.75$~mm are coloured by their doubling rate. Horizontal lines correspond to acoustic amplitudes from numericial testing and circles represent those numerical test which stimulated parametric instability growth for a given dominant perturbation mode $k$.}
\label{fig:ff-mathieu}
\end{figure*}

The stabilty of \equ{eqn:AW-mathieu} is analysed using the Floquet theory detailed in \appx{ap:floquet}. Unstable solutions to \equ{eqn:AW-mathieu} take two forms: DL instability~\cite{landau1944TheorySlowCombustion,darrieus1945PropagationDunFront} which occurs at zero and low acoustic amplitudes as well as parametric instability which requires a non-zero acoustic amplitude and yields an excited flame motion which is subharmonic (i.e.~has half the acoustic frequency). We plot in \fig{fig:ff-mathieu} the regions of intrinsic DL flame instability and parametric flame instability for given wavenumber perturbation $k$ and dimensional acoustic amplitude $p_{\rm{amp}}$. The dimensional growth rate of the DL region stays roughly constant between all three cases as this is a non-acoustic phenomenon. Dimensional analysis shows that growth rates in the parametric region grow proportionally with frequency. Values of $k$ are a parameter of the problem, so wavenumber modes are decoupled and the resulting parametric flame structures are restricted to a finite sum of sinusoids (unlike the MS equation, which explicitly couples different wavenumber modes). For flames which are parametrically unstable, the expected flame perturbation corresponds to the wavenumber $k$ with the largest doubling rate at that acoustic amplitude $p_{\rm{amp}}$. Hence, we expect $k = 1$, 2 and 3 responses for $f = 150$, 600 and 1500~Hz acoustic forcing, respectively. The vertical gap between the DL and parametric regions in \fig{fig:ff-mathieu}a, b, and c correspond to flames which reach limit cycle due to primary instability saturation and require further growth of the acoustic mode before the parametric instability is triggered. In our case, we avoid this regime by injecting acoustic waves from the outflow.

The horizontal lines in \fig{fig:ff-mathieu} correspond to a single simulation performed with acoustic waves injected at frequency $f$ at that amplitude $p_{\rm{amp}}$. Circles are placed along this line at the wavenumber corresponding to perturbations due to parametric instability, if the instability is triggered. In all three cases, the dominant wavenumber is the one expected from the above analysis. In the cases of the $f = 150$ and 600~Hz acoustics, the parametric instability threshold is observed at the correct amplitudes. This is not true for $f = 1500$~Hz: we believe this to be a consequence of the higher order flame thickness terms in the Mathieu equation \equ{eqn:AW-mathieu} having been removed. Hence, the higher wavenumbers are more likely to render quantitatively inaccurate predictions.

\begin{figure*}
\centering
\begin{subfigure}{0.99\textwidth}
\centering
\includegraphics[width=0.9\textwidth]{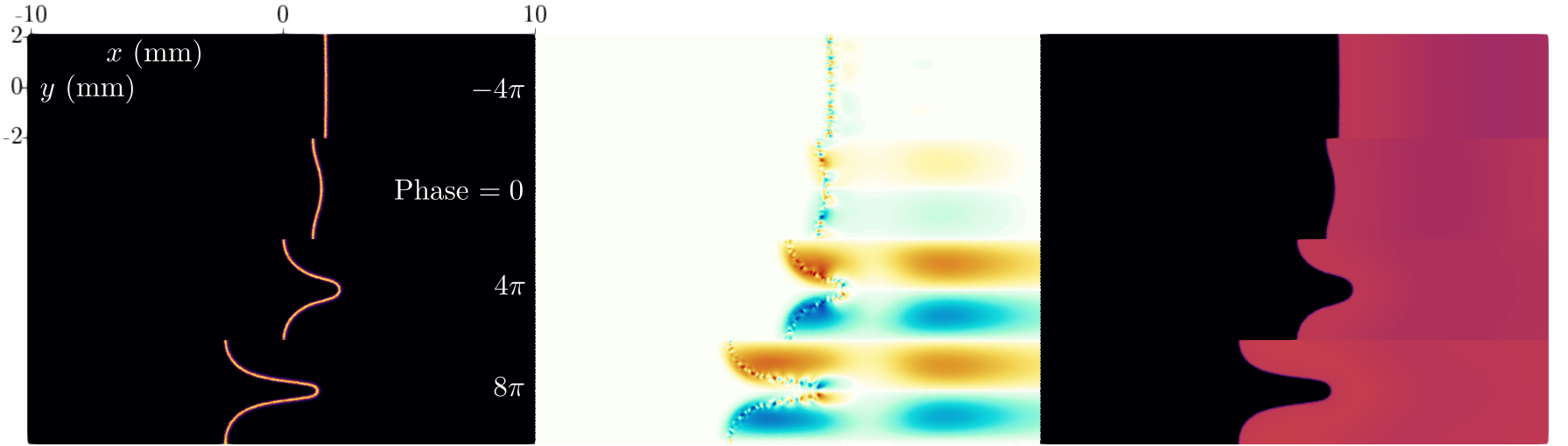}
\end{subfigure}
\vspace{0.5em}

\begin{subfigure}{0.99\textwidth}
\centering
\includegraphics[width=0.9\textwidth]{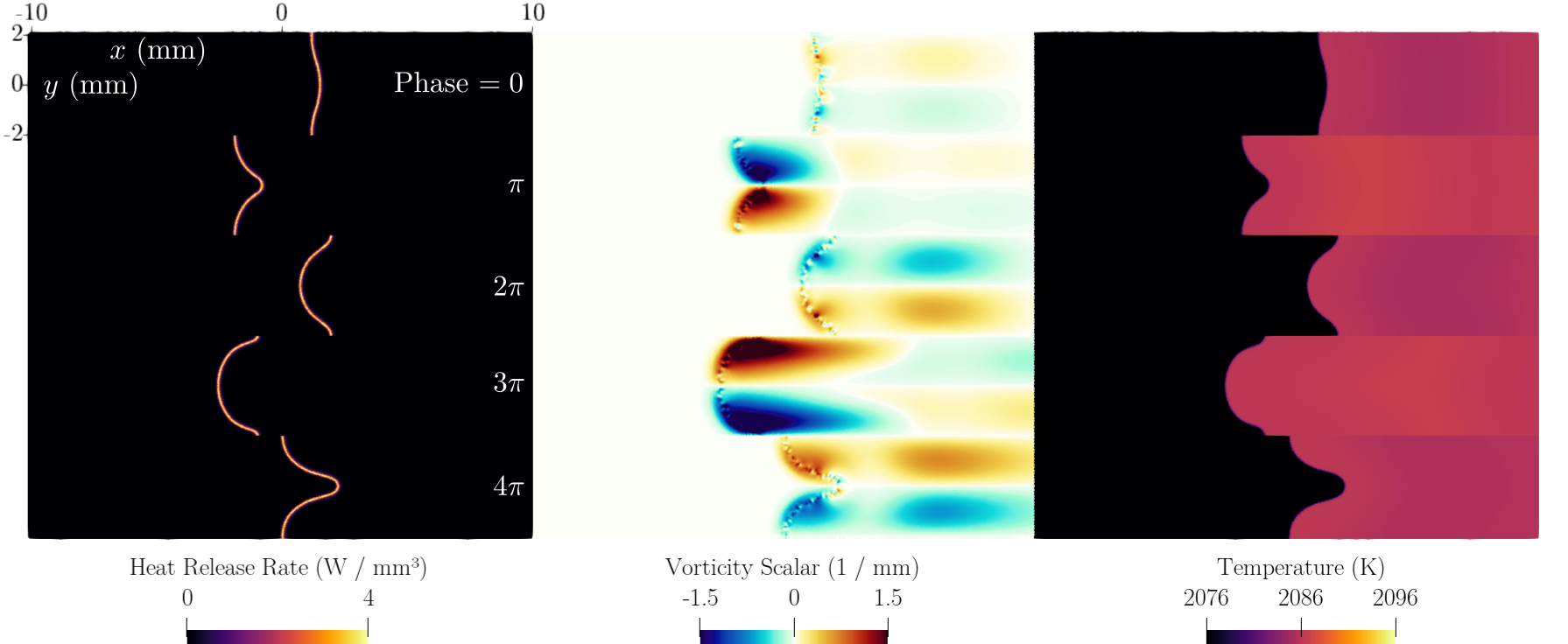}
\end{subfigure}
\caption{Parametric flame structures in the DNS region for $f = 150$~Hz. See supplementary material for video of flame evolution.}
\label{fig:ff-sec-fields_150Hz}
\end{figure*}

\begin{figure*}
\centering
\begin{subfigure}{0.99\textwidth}
\centering
\includegraphics[width=0.9\textwidth]{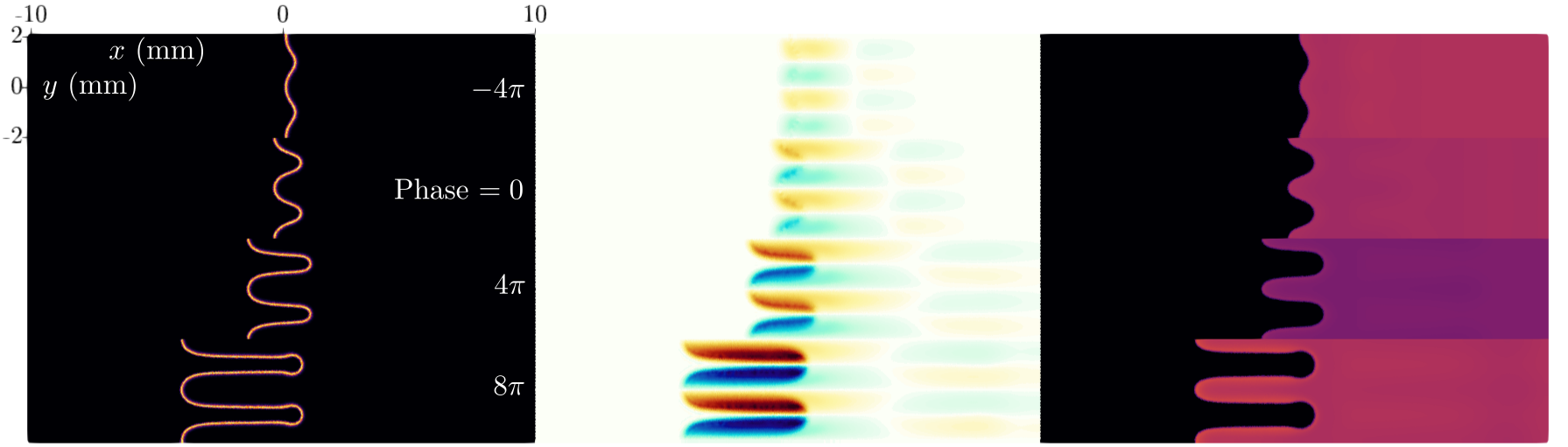}
\end{subfigure}
\vspace{0.5em}

\begin{subfigure}{0.99\textwidth}
\centering
\includegraphics[width=0.9\textwidth]{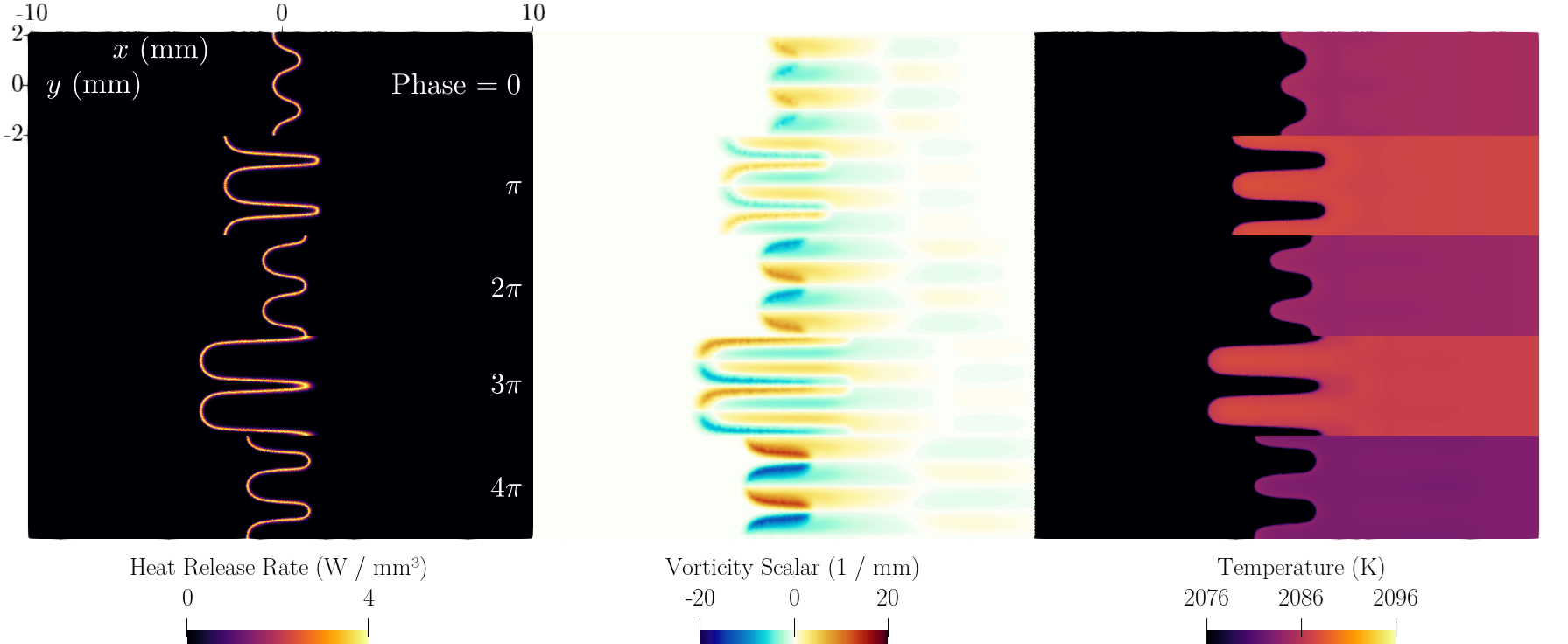}
\end{subfigure}
\caption{Parametric flame structures in the DNS region for $f = 600$~Hz. See supplementary material for video of flame evolution.}
\label{fig:ff-sec-fields_600Hz}
\end{figure*}

\begin{figure*}
\centering
\begin{subfigure}{0.99\textwidth}
\centering
\includegraphics[width=0.9\textwidth]{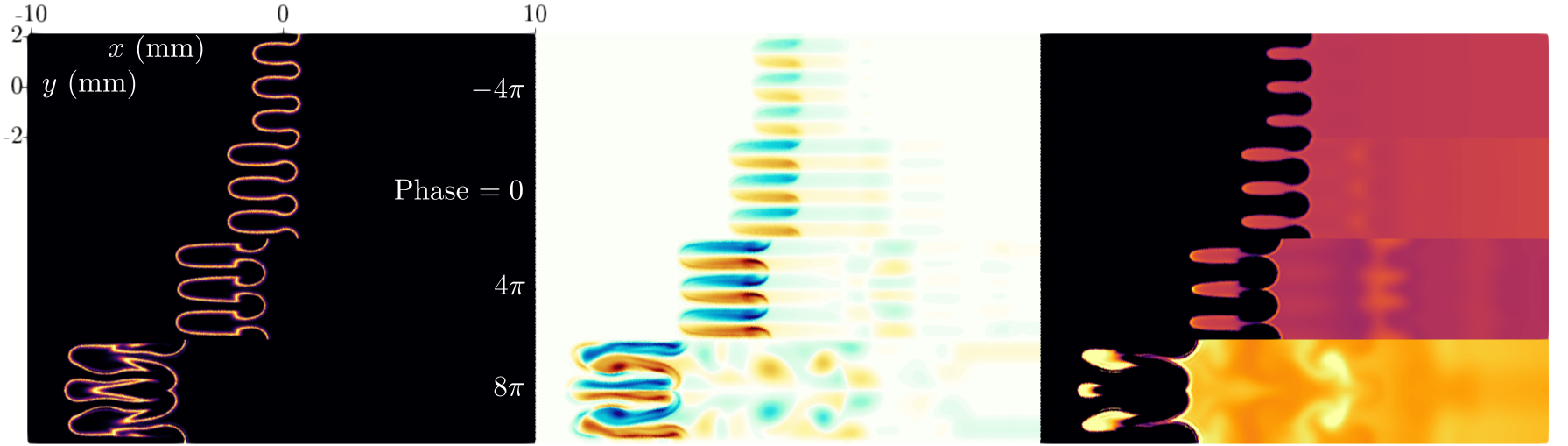}
\end{subfigure}
\vspace{0.5em}

\begin{subfigure}{0.99\textwidth}
\centering
\includegraphics[width=0.9\textwidth]{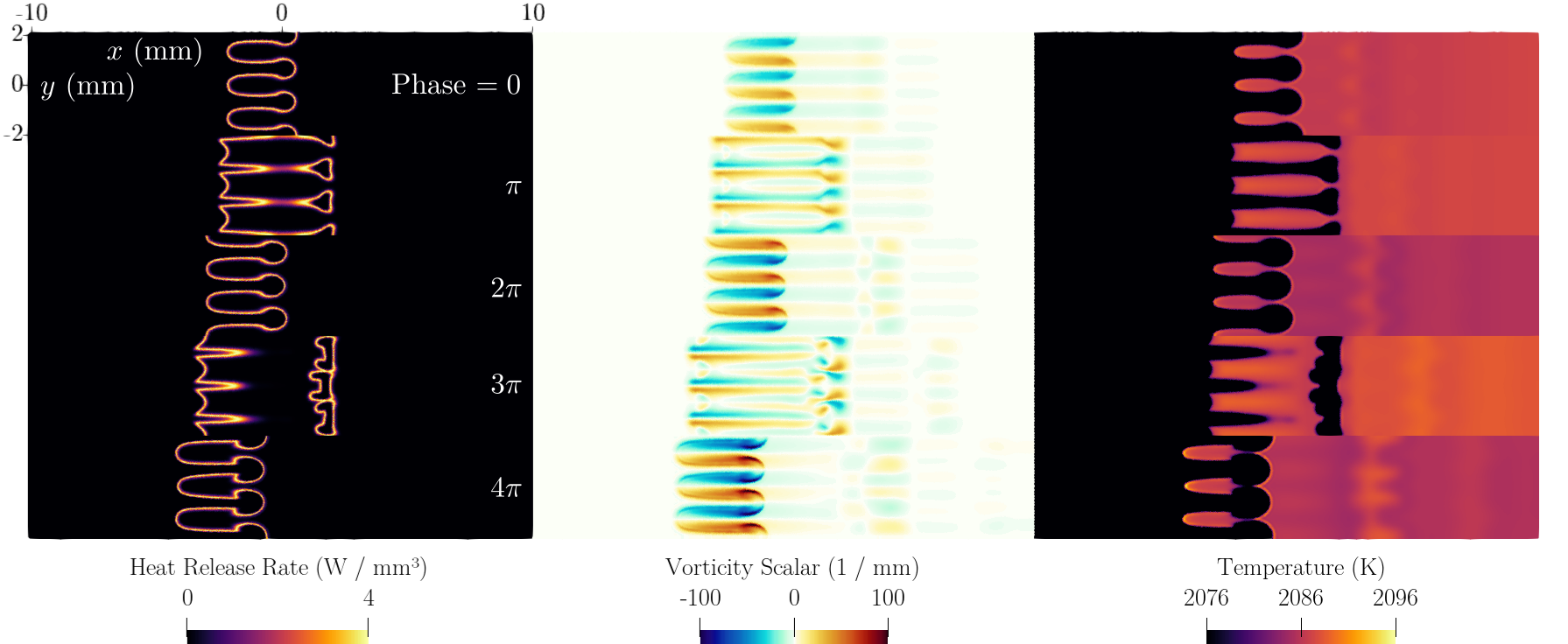}
\end{subfigure}
\caption{Parametric flame structures in the DNS region for $f = 1500$~Hz. See supplementary material for video of flame evolution.}
\label{fig:ff-sec-fields_1500Hz}
\end{figure*}

\fig{fig:ff-sec-fields_150Hz}, \fig{fig:ff-sec-fields_600Hz} and \fig{fig:ff-sec-fields_1500Hz} show how parametric instability develops over many subharmonic oscillations for $f = 150$, 600 and 1500~Hz with $L_{\rm{tube}} = 100$, 25, 10~cm, respectively. Video of flame dynamics are given in the supplementary material for this work. In each case, we choose the simulation which is parametrically unstable at the lowest acoustic amplitude. The perturbations occur independently of discretisation length scale. In all cases, the parametric instability manifests as sinusoidal perturbations to the flame to begin with. This validates the assumption in the Mathieu equation model that wavenumber modes are decoupled. As the instability develops, a flame cusping forms: the procession of cold reactant toward the hot products form flame structures with higher curvature than the flame structures at the recession of hot product into the cold reactant. In \fig{fig:ff-sec-fields_600Hz} and \fig{fig:ff-sec-fields_1500Hz} the flame structures stop being represented by a function $F = F(t, y)$ after only a few subharmonic oscillations. At this point, the Mathieu equation model of the flame response stops being valid. The subharmonically oscillating fingers present in these cases are similar to those present at the fluid interface in the RT instability, where we take the body force equal to the surrounding acoustic field. Although \fig{fig:ff-sec-fields_150Hz} demonstrates a non-cellular parametric response at $L_{\rm{tube}} = 100$~cm, this instability is still essentially cellular as increasing the dimensional width of the domain would allow for more cells to form at the wavelength we observe in this case. The DL vortex sheet~(described in e.g.~\cite{matalon2018DarrieusLandauInstability}) is also visible in \fig{fig:ff-sec-fields_150Hz} as the larger flame structures result in lower overall vorticity amplitudes. Overall temperature variations remain low due to the unity Lewis number. Flow becomes more extreme as the instability progresses. We do not observe the same oscillation every acoustic period between tulip and fingered flames observed in prior numerical and experimental literature (e.g.~\cite{jun2023ParametricInstabilityPropagating,delfin2024ThermoacousticParametricInstability}) due to the absence of walls, which encourage tulip flames~\cite{ponizy2014TulipFlameMechanism}. Although the associated flow velocities increase greatly under the secondary instability, turbulence does not develop in the simulations. This is likely due to the lack of time available for the hydrodynamic instabilities to develop before the flames reach the DNS inflow.




\section{Conclusions} \label{sec:6}
The Acoustic Delay Characteristic Boundary Condition (ADCBC) method has been introduced. The DNS domain is truncated to contain only the flame and surrounding hydrodynamics, rather than the full acoustic domain. This is implemented as a modifcation to the popular Navier-Stokes Characteristic Boundary Condition method~\cite{poinsot1992BoundaryConditionsDirect}. By storing the change in acoustic waves leaving the domain at the in- and outflows, we can model the truncated up- and downstream sections of the tube's one-dimensional acoustics with a simple time delay before their reentry into the domain. The modified in- and outflow boundary conditions were applied to DNS of several inert and idealised reactive flows for validation. Other discretisation methods besides that employed in this paper may be used alongside the ADCBC method, provided the NSCBC formulation is used at the boundaries. First, an acoustic travelling wave test case was used in a short domain with ADCBC inflow and outflows. This reproduced the structure of the acoustic signal after one and many reflections via the in- and outflows. An acoustic standing wave test case was used in the same domain, where the wave structure simulated over many acoustic periods was maintained.

The Averaged Proportional and Integral Linear Relaxation (APILR) method has also been introduced for these boundary conditions to suppress numerical velocity and pressure drift. Moving averages act as a lowpass filter for frequencies corresponding to the characteristic tube frequency and above. An integral term was introduced as well as the proportional term present in the Classic Linear Relaxation (CLR) method~\cite{rudy1980NonreflectingOutflowBoundary}. This integral term suppresses non-zero equilibrium drift. When drift control was used, once the flame flattens, the increased acoustic amplitude does not result in a drift in the average inflow velocity. This ensures the flat flame remains stationary in a counterflow equal to the laminar flame speed.

An idealised premixed flame in a 100~cm long closed-open tube was simlated in a 2~cm long by 4~mm wide truncated DNS domain using the ADCBC and APILR methods. We observe primary thermoacoustic instability as the flame flattens due to the growth of the fundamental mode and first harmonic. At this point, primary instability growth rate saturates due to nonlinear effects. This corroborates results found throughout the literature and greatly reduces the computational expense of these simulations. The APILR drift control terms effectively contain the flame in the DNS domain and respond as expected to the primary thermoacoustic instability. The same flame is then simulated in 10, 25 and 100~cm long tubes with acoustic forcing at the fundamental acoustic mode: cellular parametric instability is observed in all cases with the flames having three, two or one wavelength along the domain width, respectively. These cells oscillate subharmonically, resulting in periodic flame fingering in the 10 and 25~cm long tubes due to the unsteady Rayleigh-Taylor instability. The flame's wavenumber response corresponds to that predicted by the Mathieu equation derived from~\cite{searby1991ParametricAcousticInstability} provided we choose non-dimensional width parameter $\g$ such that the theoretical planar stability limit at critical value $\g^c = 2$ agrees with stability limit found from numerical trial-and-error.

The boundary conditions can just as easily be applied to more realistic flames involving e.g.~multiple species, mixture averaged and temperature-dependent transport properties and multiple step chemistry. This presents an interesting avenue for further research using these boundary conditions. In the case of thermodiffusive flames, the inflow drift control scheme could be used to follow a target velocity which tracks the flame speed. Characteristic boundary conditions under the NSCBC formulation have also been used in the literature for turbulent flows. This would enable study of the combined interaction between thermoacoustic, turbulent and thermodiffisive instabilities. Futhermore, signal processing methods may be used on stored acoustic values to model a frequency dependence in the reflection coefficient (c.f. impedance boundary conditions). In all the above applications, the ADCBC and APILR schemes could offer significant savings in computational cost.

\bibliography{references}

\appendix

\section{Calculation of Acoustic Eigenmodes} \label{ap:eigenmodes}

This appendix follows a simplified form of the analysis in~\cite{clavin1990OnedimensionalVibratoryInstability} to leading order in Mach number. All variables have been non-dimensionalised by: tube length $\dimvar{L}_{\rm{tube}}$ for space such that $x \in [0, 1]$, velocity by laminar flame speed $\dimvar{S}_L$, time by characteristic acoustic time $\dimvar{L}_{\rm{tube}} / \dimvar{c}_{\rm{U}}$, density by upstream density $\dimvar{\r}_{\rm{U}}$, pressure by upstream acoustic pressure $\dimvar{\r}_{\rm{U}} \dimvar{c}_{\rm{U}} \dimvar{S}_L$ and temperature by upstream temperature $\dimvar{T}_{\rm{U}}$. Asterisks represent the dimensional variables in this section. Hence, far upstream temperatures are $T=1$ and far downstream temperatures are $T=1 + q$. By the dimensional relation $\dimvar{\r} (\dimvar{c})^2 = \g \, \dimvar{p}$ we have that $c_{\rm{D}}^2 = (1 + q) c_{\rm{U}}^2$. The linearised equations for acoustic pressure and velocity in the closed-open tube's primary axis are the wave equations:
\begin{subequations} \label{eqn:wave_eqn}
\begin{align}
\pdv[2]{u_a}{t} - T \pdv[2]{u_a}{x} &= 0 \label{eqn:wave_eqn_u} \\
\pdv[2]{p_a}{t} - \pdv{x} \left( T \pdv{p_a}{x} \right) &= 0. \label{eqn:wave_eqn_p}
\end{align}
\end{subequations}
Acoustic pressures and velocities are related by $\overline{\r} \, \partial u_a / \partial t = \partial p_a / \partial x$. We have made the assumption of a non-convected flow $\overline{u} = 0$ and constant background pressure over the flame $\overline{p}_{\rm{U}} = \overline{p}_{\rm{D}}$ which is a good approximation for low speed flames. Boundary conditions $u_a(t, x = 0) = 0$ and $p_a(t, x = 1) = 0$ model the closed-open tube.

\begin{figure*}[t]
\centering
\includegraphics[scale=0.34]{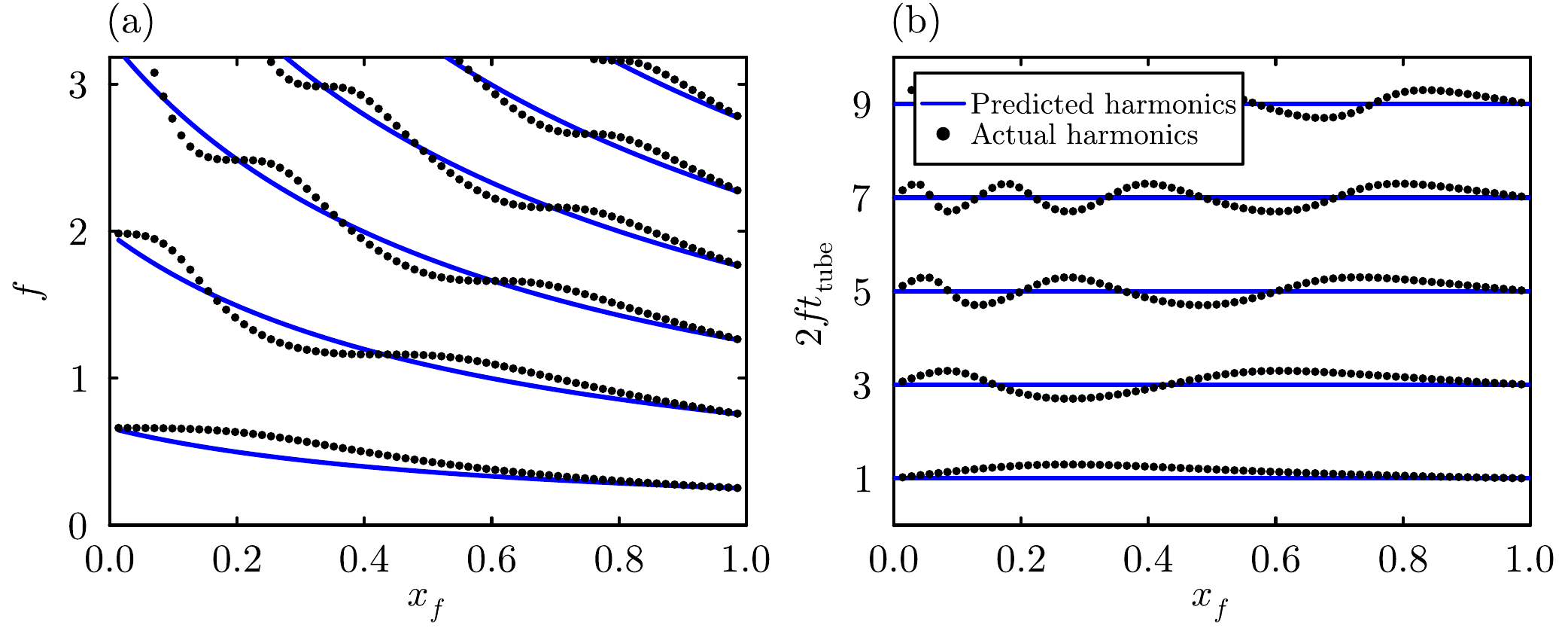}
\caption{Predicted harmonics and true harmonics for $q = 6$.}
\label{fig:harmonics}
\end{figure*}

With the flame located at $x_f \in (0, 1)$, each field is separated into up- and downstream domains: $T_{\rm{U}} = 1$, $T_{\rm{D}} = 1 + q$, $u_a(t, x_f^-) = u_{a, \rm{U}}(t, x_f^-)$, $u_a(t, x_f^+) = u_{a, \rm{D}}(t, x_f^+)$ and so on. The wave equations \equ{eqn:wave_eqn} describe the propagation on either side of the flame. To enforce continuity of mass flux we prescribe jump conditions over the flame as:
\begin{subequations} \label{eqn:flame-BCs}
\begin{align}
p_{a, \rm{U}}(t, x_f^-) &= p_{a, \rm{D}}(t, x_f^+), \\
u_{a, \rm{U}}(t, x_f^-) &= (1 + q) \, u_{a, \rm{D}}(t, x_f^+). \label{eqn:flame-BCs-u}
\end{align}
\end{subequations}
The low-Mach assumption is responsible for \equ{eqn:flame-BCs-u} as the corresponding jump condition in~\cite{clavin1990OnedimensionalVibratoryInstability} is first-order in Mach number. This implicitly contains information on how waves are transmitted and reflected over the density jump. Further, we presume the flame behaves as a passive interface, so radiation of acoustic waves due to a flame transfer function are neglected. Note that if $q = 0$ or $x_f = 0, 1$, we end up in the situation of isothermal acoustics, which have the classical sinusoidal eigenmode solutions. To find non-isothermal eigenmodes, we use the time-harmonic factor $u_a, p_a \propto \exp(i 2\p\!f t)$ where $f \in \mathbb{R}$. Note that the wave equation model involves no damping terms so an imaginary component of $f$ is not required. Using the time-harmonic assumption, we define the function $\f(x)$ such that $p_a(t, x) = \f(x) \exp(-i 2\p\!f t)$, so the second order PDE in time and space becomes a second order ODE in space:
\begin{subequations}
\begin{gather} \label{eqn:BVP_phi}
\f'' + \frac{(2\p\!f)^2}{T} \f = 0, \\
\f_{\rm{U}}'(0)=0,
\quad
\f_{\rm{D}}(1)=0,
\quad
\f_{\rm{U}} (x_f^-) = \f_{\rm{D}}(x_f^+),
\quad
(1 + q) \, \f_{\rm{U}}'(x_f^-) = \f_{\rm{D}}'(x_f^+).
\end{gather}
\end{subequations}
The complementary functions of this equation are defined on either side of $x_f$ by solutions to the complex harmonic oscillator:
\begin{equation}
\f_{\rm{U/D}}(x) = a_{\rm{U/D}} \exp\left(- i \frac{2\p\!f}{\sqrt{T_{\rm{U/D}}}} x\right) + b_{\rm{U/D}} \exp\left(i \frac{2\p\!f}{\sqrt{T_{\rm{U/D}}}} x\right),
\end{equation}
where $a, b \in \bb{C}$ represent an extra four degrees of freedom for each side of the flame for each eigenmode. Satisfying the boundary constraints requires finding the non-trivial solutions to the linear equation:
\begin{subequations}
\begin{equation}
\begin{pmatrix}
\f_{\rm{U}}'(0) \\
\f_{\rm{D}}(1)  \\
\f_{\rm{D}}(x_f^+) - \f_{\rm{U}} (x_f^-)  \\
\f_{\rm{D}}'(x_f^+) - (1 + q) \, \f_{\rm{U}}'(x_f^-)
\end{pmatrix}
= A \begin{pmatrix}
a_{\rm{U}} \\
b_{\rm{U}} \\
a_{\rm{D}} \\
b_{\rm{D}}
\end{pmatrix} = \begin{pmatrix} 0 \\ 0 \\ 0 \\ 0 \end{pmatrix}.
\end{equation}
\end{subequations}
These non-trivial solutions occur only when $\det(A) = 0$ and is solved via Newton's method. \fig{fig:harmonics} shows these harmonics as well as the expected harmonics for a range of flame positions with $q = 6$. The true harmonics oscillate about the predicted one-quarter, three-quarter etc.~modes of an equivalent homogeneous tube of the same acoustic period.

\section{Mathieu Equation Stability Analysis} \label{ap:floquet}
This analysis follows~\cite{assier2014LinearWeaklyNonlinear} where we omit dependence on the parameter $k$ for clarity. By linearity we may write \equ{eqn:AW-mathieu} as a system of ODEs for $\vb{x} := (\ftvar{F}, \dd \ftvar{F} / \dd t )\tran$:
\begin{subequations}
\begin{gather} \label{eqn:mathieu-vec}
\dv{t}\vb{x}(t) = M(t) \vb{x}(t), \\
\vb{x}(0) = ( \ftvar{F}(0), \dd \ftvar{F} / \dd t \/ (0) )\tran
\end{gather}
\end{subequations}
To analyse the stability of \equ{eqn:mathieu-vec}, we repose the equation in matrix form:
\begin{equation} \label{eqn:mathieu-mat}
\dv{t}X(t) = M(t) X(t),
\end{equation}
where $X(t) = (\vb{x}_1(t), \vb{x}_2(t))$, along with some suitable initial conditions, where the solutions $\vb{x}_{1,2}(t)$ are any two linearly independent solutions to \equ{eqn:mathieu-vec}. Since $M(t)$ is periodic with period $t_a := 1 / f$, we can write:
\begin{equation}
X(t + t_a) = C X(t)
\end{equation}
for a constant matrix $C$. Solving the eigenvalue problem for $C$, we have $AV = V\L$ where the matrix of eigenvectors $V := (\vb{v}_1, \vb{v}_2)$ and eigenvalues $\L := \rm{diag}(\l_1, \l_2)$. We use these to write solutions to the equation after $n := \lfloor t / t_a \rfloor$ periods if we start with initial conditions $X(0) = V B$ for some $B$:
\begin{equation}
X(n t_a) = C^n X(0) = V ~ \rm{diag}(\l_1^n, \l_2^n) ~ B.
\end{equation}
Hence, the stability of these initial conditions can be determined entirely by the eigenvalues of the matrix $C$. Specifically, stability is achieved when $\s := \max \{|\l_1|, |\l_2|\} < 1$ and we have instability whenever $\s > 1$. The matrix $C$ can be found by integrating in time, starting from the identity matrix $X(0) = I_2$ such that $X(t_a) = C$. For numerical time integration, we use standard fourth-order Runge-Kutta with a uniform time step $\d t = t_a / 100$.

To relate these results to DNS data, we plot heatmaps of doubling rate in units of inverse time; the solution to \equ{eqn:mathieu-vec} behaves like:
\begin{equation}
|\vb{x}(t)| \sim |\vb{x}(0)| \s^{n} \sim |\vb{x}(0)| 2^{t \bar{\o}}
\end{equation}
where we define the doubling rate $\bar{\o} = f \log_2(\s)$ such that, after redimensionalising, $\bar{\o}$ has units of Hz.

\end{document}